\documentclass[12pt]{iopart}

\RequirePackage{mathtools}

\usepackage{iopams}
\usepackage{mathrsfs}
\usepackage{amsfonts}
\usepackage{latexsym}
\usepackage{amsthm}
\usepackage{amsmath}
\usepackage{graphicx}
\usepackage{braket}
\usepackage{xcolor}
\usepackage[colorlinks,linkcolor=blue,citecolor=blue,urlcolor=blue]{hyperref}
\usepackage{subfigure}
\usepackage{enumitem}
\usepackage{scrextend}
\usepackage{multirow}

\begin{document}

\title[]{Delocalization effects, entanglement entropy and spectral collapse of boson
mixtures in a double well
}

\author{F. Lingua$^1$, G. Mazzarella$^2$, V. Penna$^1$
}
\address{$^1$
Dipartimento di Scienza Applicata e Tecnologia and U.d.R. CNISM, Politecnico di Torino, 
I-10129 Torino, Italy}
\address{$^2$ 
Dipartimento di Fisica e Astronomia Galileo Galilei and CNISM, 
Universit$\grave{a}$ di Padova, Via Marzolo 8, I-35131 Padova, Italy}
\ead{vittorio.penna@polito.it}

\begin{abstract}
We investigate the ground-state properties of a two-species condensate of interacting bosons
in a double-well potential. Each atomic species is described by a two-space-mode Bose-Hubbard
model.
The coupling of the two species is controlled by the interspecies interaction $W$. 
To analyze the ground state when $W$ is varied in both the repulsive ($W>0$) and the attractive
($W<0$) regime, we apply two different approaches. First we solve the problem numerically
i) to obtain an exact description of the ground-state structure and ii ) to characterize its correlation 
properties by studying (the appropriate extensions to the present case of) the quantum Fisher 
information, the coherence visibility and the entanglement entropy as functions of $W$. 
Then we approach analytically the description of the low-energy scenario 
by means of the Bogoliubov scheme. In this framework
the ground-state transition from delocalized to localized species (with space separation for 
$W>0$, and mixing for $W<0$) is well reproduced. 
These predictions are qualitatively corroborated by our numerical results.
We show that such a transition features a spectral collapse reflecting the dramatic change of 
the dynamical algebra of the four-mode model Hamiltonian.
%
\end{abstract}
\vspace{2pc}

\maketitle

\section{Introduction}
\label{sec1}

Over the past decade, mixtures formed by two bosonic species have raised a considerable
interest \cite{kuklov1}-\cite{BD} due to the theoretic prediction of new exotic quantum
phases. Such phases, determined by the competition of the interspecies interaction with
the onsite interactions and tunneling amplitudes of each species, include, for example,
unprecedented Mott-like states and types of superfluidity \cite{kuklov1}-\cite{isacsson},
and composite states exhibiting coexistence of different phases \cite{capogrosso2}.
Mixtures, realized by means of
either two atomic species \cite{catani} or the same species in two different internal
states \cite{gadway}, have been trapped in optical lattices formed by many potential wells
and are efficiently described in the Bose-Hubbard (BH) model framework \cite{fisher,jaksch}.
Their recent in-lab achievement has made stronger the interest for these systems.

The simplest but not trivial optical device where condensates of different atomic
species can be loaded is the trap formed by a double-well potential. This has been realized
by Albiez and co-workers \cite{doublewell-heidelberg} in 2005 by superposing a parabolic trap
with a (sinusoidal) linear potential.
This system has been widely explored within the mean-field approach to analyze
the atomic counterpart of the Josephson effect in superconductor-oxide-superconductor
junctions \cite{book-barone}, 
the dynamical stability of binary mixtures \cite{lobo,xu},
different types of self-trapping solutions \cite{satija},
the effectiveness of the space-mode description within
the Gross-Pitaevskii picture \cite{moratti1,diaz2}
and the Rabi-Josephson regime \cite{mazz-rabi}.
Further work has shown how the double-well system with two species
can be exploited to model the dynamics of spin-orbit-coupled condensates
\cite{chineseguys-so,miguel-so}, to investigate their quantum evolution 
in the Raman-coupling regime \cite{citro2} and energy-spectrum properties
\cite{wang2015}.
%
%

For several reasons, a fully-quantum analysis of few-well systems deserves a treatment too. 
First, because in the presence of sufficiently-low boson numbers only a 
BH-like formulation in terms of second-quantized space modes can provide a realistic 
description of microscopic processes in mixtures at essentially zero temperature.
Second, because even the relatively simple models
consisting of a mixture trapped in two wells feature complex dynamical behaviors
owing to the strong nonlinearity of the interaction terms. This character affects,
not only high-energy but also weakly-excited states including the ground state. Then,
at very low temperatures, it is interesting to revisit nonlinear behaviors, first explored
classically, in a fully-quantum environment.
Finally, various well-established methods, both numerical and theoretical,
seem to be applicable to study quantum models describing a mixture in two wells.
Despite such circumstances, a rather small attention has been focused on quantum aspects of these
systems \cite{naddeo2010}-\cite{Zi}.

This work aims to explore the low-energy properties of a
two-species condensate confined in a double-well potential. The model
describing this mixture is therefore a two-species Bose-Hubbard Hamiltonian
defined on a two-site lattice (dimer) that, in addition to single-component interwell
tunneling and onsite boson-boson interactions, also incorporates the density-density
interaction $W$ between atoms of different species. For brevity we call this model
the two-species dimer (TSD).

The first aspect we consider is the numerical diagonalization of the TSD Hamiltonian
in the zero-temperature regime.
We find that, by keeping fixed the intraspecies interaction $U$, variations of
interspecies interaction $W$ produce significant changes in the structure of the ground state.
On the repulsive side, increasing $W$ drives the ground state from
the uniform state (delocalization regime with mixed species) to a symmetric superposition
of states exhibiting a macroscopic space separation (demixing) of the two species,
passing through an intermediate configuration where the delocalized state and the symmetric
demixed state coexist.
When the interspecies interaction is attractive, instead, the uniform ground state
``evolves" to a configuration in which atoms of different components tend to macroscopically
populate the same well (localization regime with mixed species).

Studying appropriate extensions to the present case of the Fisher information
\cite{braunstein,pezze}, the coherence visibility $\alpha$ \cite{stringari,anglin,anna}
and the entanglement entropy $S$ \cite{bwae} in terms of interaction $W$
allows us to characterize the ground state from the correlations point of view.
In particular, we find that the maxima of $\alpha$ and $S$ correspond to the on-set of
a ground state in which a delocalized component and two symmetric localized states coexist.
These numerical results improve our insight on how the ground-state structure
depends on $W$.

The information obtained from the numerical determination of the ground-state
configurations is then used to approach analytically the investigation of the
low-energy regimes within the Bogoliubov approximation.
In this framework we diagonalize the TSD Hamiltonian, in the strong-interaction ($|W|>U$)
and in the weak-interaction ($|W|<U$) regime, 
by identifying weakly-occupied modes
in the space-mode picture and in the momentum-mode picture, respectively. We reconstruct
the low-energy spectrum, comparing the latter with the numerical eigenspectrum, and 
provide an approximate form of weakly-excited states.
The demixing condition written in terms of the model parameters is derived analytically.
We shed light on the spectral collapse occurring when the localization-delocalization 
transition of the species takes place, which causes the change of the energy spectrum from 
a discrete form to an almost continuous one \cite{vittorio2013}-\cite{felicetti}.
The corresponding change of the algebraic structure of the Hamiltonian is also discussed.

\section{The 2-species dimer Hamiltonian}
\label{sec2}

We consider two interacting
bosonic gases (denoted below by $a$ and $b$) at zero temperature. Each species
is confined in a three-dimensional trapping potential $V_{trap,c}({\bf r})$
($c=a,b$) achieved by superimposing an isotropic harmonic confinement in the
transverse radial ($x$-$y$) plane and a symmetric double-well $V_{DW,c}(z)$ potential
in the axial direction $z$. The resulting potential is
$V_{trap,c}({\bf r})=m_c\omega_{c}^2(x^2+y^2)/2+V_{DW,c}(z)$,
where $m_c$ is the boson mass of the $c$th component,
$\omega_c$ the trapping frequency in the radial plane felt by the species $c$, and
quantities $|m_a-m_b|$, $|\omega_{a}-\omega_{b}|$ are assumed to be sufficiently small.
Due to the strong radial harmonic confinement we treat the system as quasi-one-dimensional.
Moreover, if the energy per particle in the axial direction is much smaller than
the transverse level spacing $\hbar \omega_c$, then bosons can be assumed to stay
in the ground state of the transverse harmonic oscillator.
The model describing this system can be derived from the bosonic-field Hamiltonian
by implementing the space-mode approximation (see, for example, \cite{moratti2,milb}).
By expanding the field operators in the basis of the single-particle wave functions
localized in each well, one achieves the two-species BH Hamiltonian
%
\begin{equation}
\label{bhtwo}
\hat{H} ={\sum}_{c= a, b}\hat{H}_{c}+\hat{H}_{ab}
\end{equation}
with
$$
\hat{H}_{c}=
\frac{U_c}{2}\Bigl [
\hat{c}_{L}^{\dagger}\hat{c}_{L}^{\dagger}\hat{c}_{L}\hat{c}_{L}
+
\hat{c}_{R}^{\dagger}\hat{c}_{R}^{\dagger}\hat{c}_{R}\hat{c}_{R} \Bigr ]
-J_c \big(\hat{c}_{L}^{\dagger}\hat{c}_{R}+\hat{c}_{R}^{\dagger}\hat{c}_{L} \big ),
$$
and
$$
\hat{H}_{ab}=
W \big (  \hat{a}^{\dagger}_{L} \hat{a}_{L} \hat{b}^{\dagger}_{L}  \hat{b}_{L}
+
\hat{a}^{\dagger}_{R} \hat{a}_{R} \hat{b}^{\dagger}_{R} \hat{b}_{R}\big) \; .
$$
The operator $\hat{c}_v= \hat{a}_v (\hat{b}_v)$
annihilates a boson of species $a$ ($b$) in the $v$th well
($v=L,R$, with $L$ ($R$) representing the left (right) well), $U_c$ is the amplitude
of the intraspecies onsite interaction,
$J_c$ is the tunneling amplitude, and $W$ is the interspecies interaction
amplitude between bosons in the same well. 
Boson operators $a_v$, $a_v^\dagger$, $b_v$ and $b_v^\dagger$,
in addition to $[a_v, a^\dagger_u] = \delta_{uv}= [b_v, b^\dagger_u]$
with $u = L,R$, satisfy the commutators $[a_v, b_u]=0$, and $[a_v, b^\dagger_u]=0$. 
For each species, the total boson
numbers ${\hat N}_c$ commute with ${\hat H}$ implying that
$N_a = N_{aL}+N_{aR}$ and $N_b = N_{bL}+N_{bR}$ are conserved quantities.
To further simplify the model we assume species $a$ and $b$
with the same mass, the same intraspecies s-wave scattering length, and experiencing
identical double-well and harmonic confinements. In the following, we thus set
$J_a=J_b \equiv J$ and $U_a=U_b \equiv U$, that is, the components $a$ and $b$
have the same hopping parameter and intraspecies interaction. We shall relax such
assumptions whenever is possible.

\section{Quantum analysis}
\label{qa}

We solve the eigenvalue equation associated to Hamiltonian (\ref{bhtwo})
\begin{equation}
\label{eigenproblem}
\hat{H}\,|\psi\rangle_n=E_n\,|\psi\rangle_n\;
\end{equation}
for fixed numbers $N_a$ and $N_b$ of bosons of species $a$ and $b$, respectively.
In this case the Hamiltonian can be represented by a $M \times M$ matrix with $M=(N_a+1)(N_b+1)$
in the basis $|i, j\rangle_L \otimes |N_a-i,N_b-j\rangle_R$ with $i\in [0,N_a]$ and $j \in [0,N_b]$.
For each eigenvalue $E_n$, with $n=1,2,...,M$, the corresponding eigenstate $|\psi\rangle_n$
has the form
\begin{equation}
\label{superposition}
|\psi\rangle_n=\sum_{i=0}^{N_a}\,\sum_{j=0}^{N_b} 
C_n (i,j)
|i, j\rangle_L |N_a-i,N_b-j\rangle_R
\; ,
\end{equation}
where we assume --without loss of generality-- that the coefficients $C_n ({i,j})$ are real.
In the ket $|i,j\rangle_L$,
$i$ ($j$) is the number of bosons of species $a$ ($b$) in the left well, while
in $|N_a-i,N_b-j\rangle_R$,  $N_a-i$ ($N_b-j$) is the number of bosons of species
$a$ ($b$) in the right well.
Since we focus our attention on the ground state ($n=1$),
we simplify the notation by setting $C_1 ({i,j}) \equiv C({i,j})$.
When both the intra- and the interspecies interactions are zero, the ground state of
Hamiltonian (\ref{bhtwo}) is given by the product of two atomic (or SU(2))
coherent states (see, e. g., \cite{gilm}, \cite{jpa41})
\begin{equation}
\label{coherentdouble}
|ACS \rangle =  {1\over \sqrt{N_a! N_b!}}
\left ( \frac{{\hat a}_{L}^{\dagger} + {\hat a}_{R}^{\dagger}}{\sqrt 2}
\right )^{N_{a}}
\left ( \frac{{\hat b}_{L}^{\dagger} + {\hat b}_{R}^{\dagger}}{\sqrt 2}
\right )^{N_{b}}
|0,0\rangle_L |0,0\rangle_R
\; ,
\end{equation}
where $|0,0\rangle_L |0,0\rangle_R$ is the state with no bosons.
Coherent states of this form usually describe the superfluid phase for large $J_c/U_c$
where bosons are uniformly distributed on the lattice and thus totally delocalized.

\begin{figure}[h]
\begin{center}
\begin{tabular}{cc}
\includegraphics[
clip,width=0.3\textwidth ]{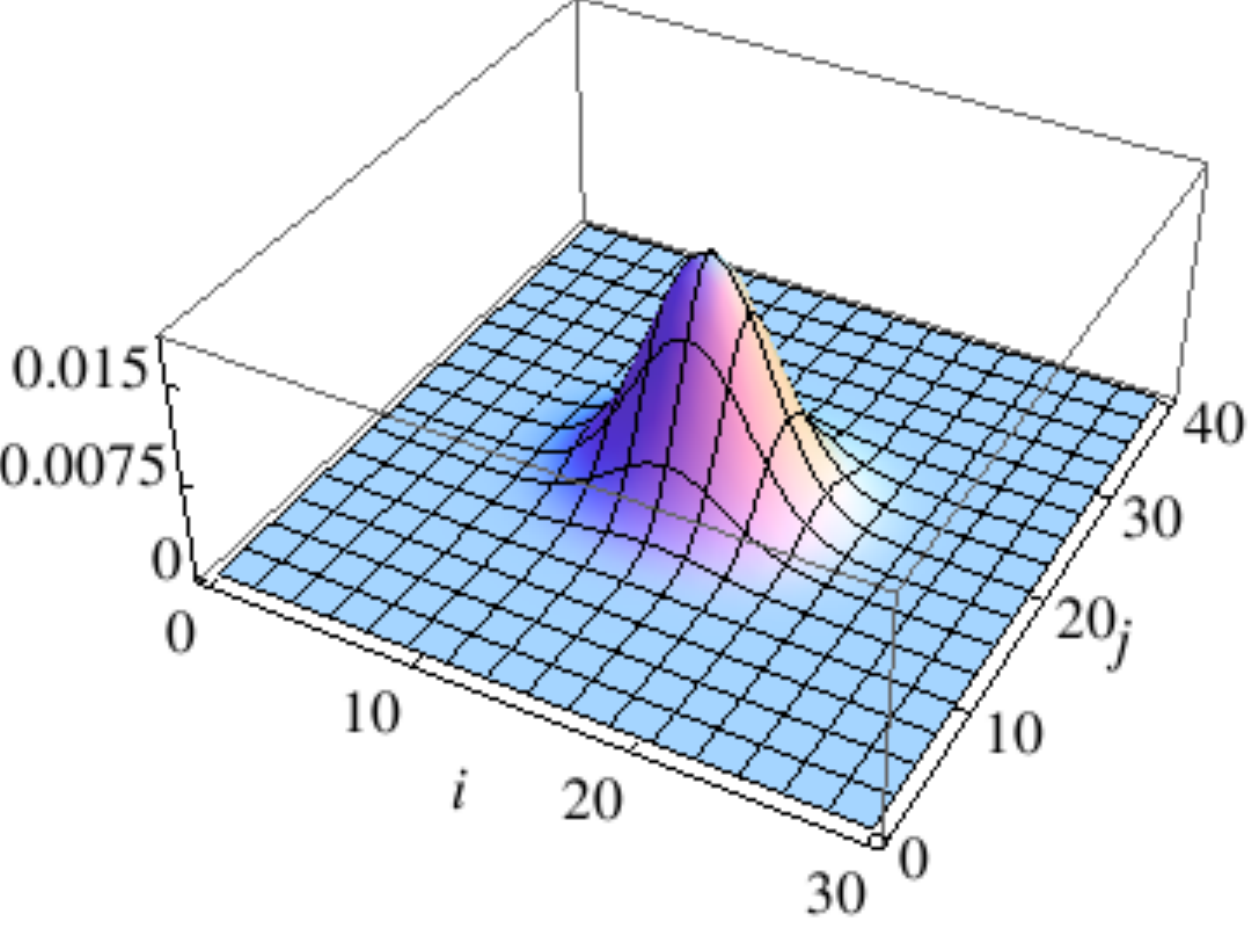}
&
\includegraphics[
clip,width=0.3\textwidth ]{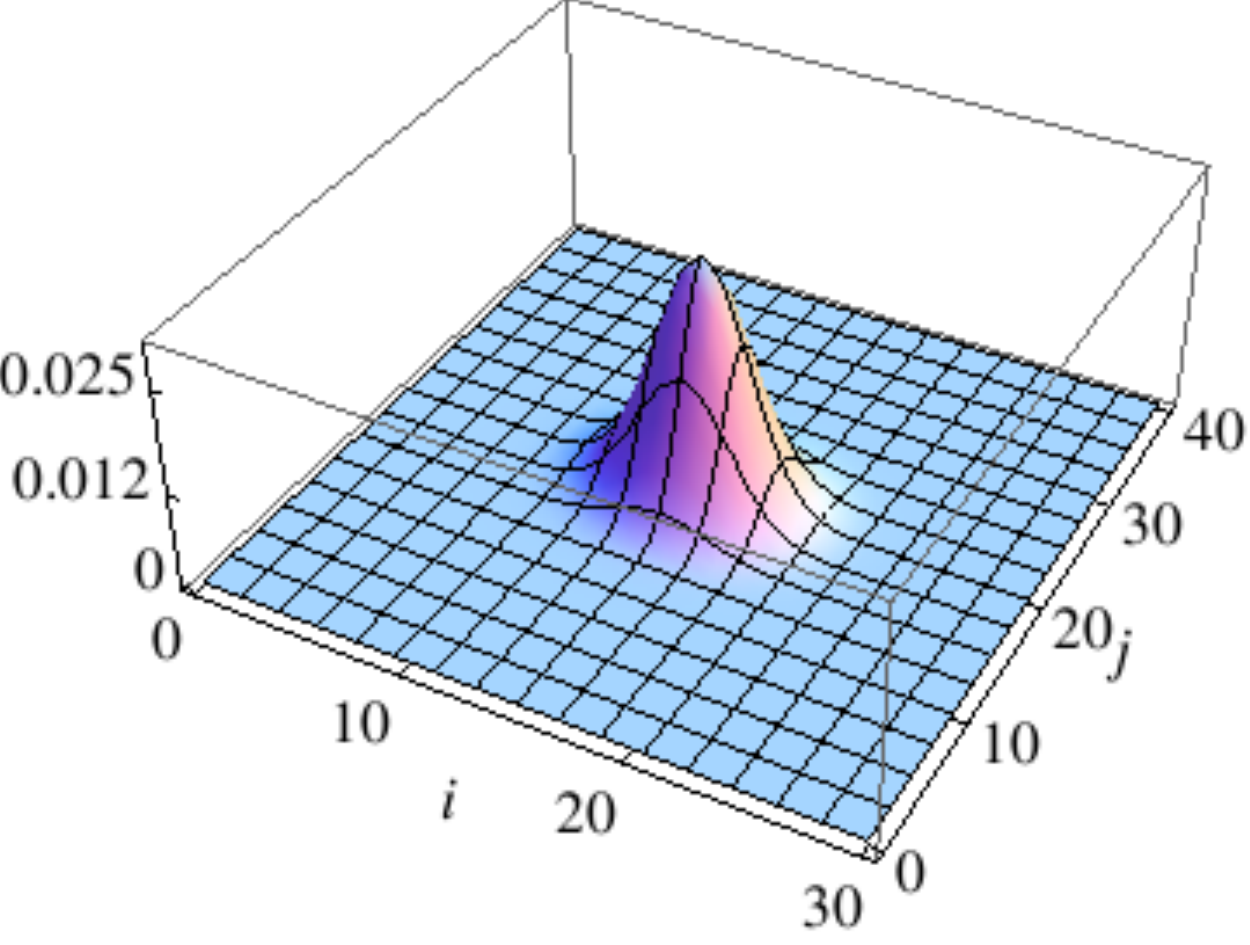}
\end{tabular}
\caption{(Color online) 
Ground-state coefficients $|C({i,j})|^2$ vs $i$ (left occupation numbers of species $a$) 
and $j$ (left occupation number of species $b$) for $W=0$, $U=0$ (left panel), $U=0.1$ (right panel)  
and boson numbers $N_a=30$, $N_b=40$.
Energies in units of $J$.}
\label{fig1n}
\end{center}
\end{figure}

\begin{figure}[h]
\begin{center}
\begin{tabular}{ccc}
\includegraphics[
clip,width=0.3\textwidth ]{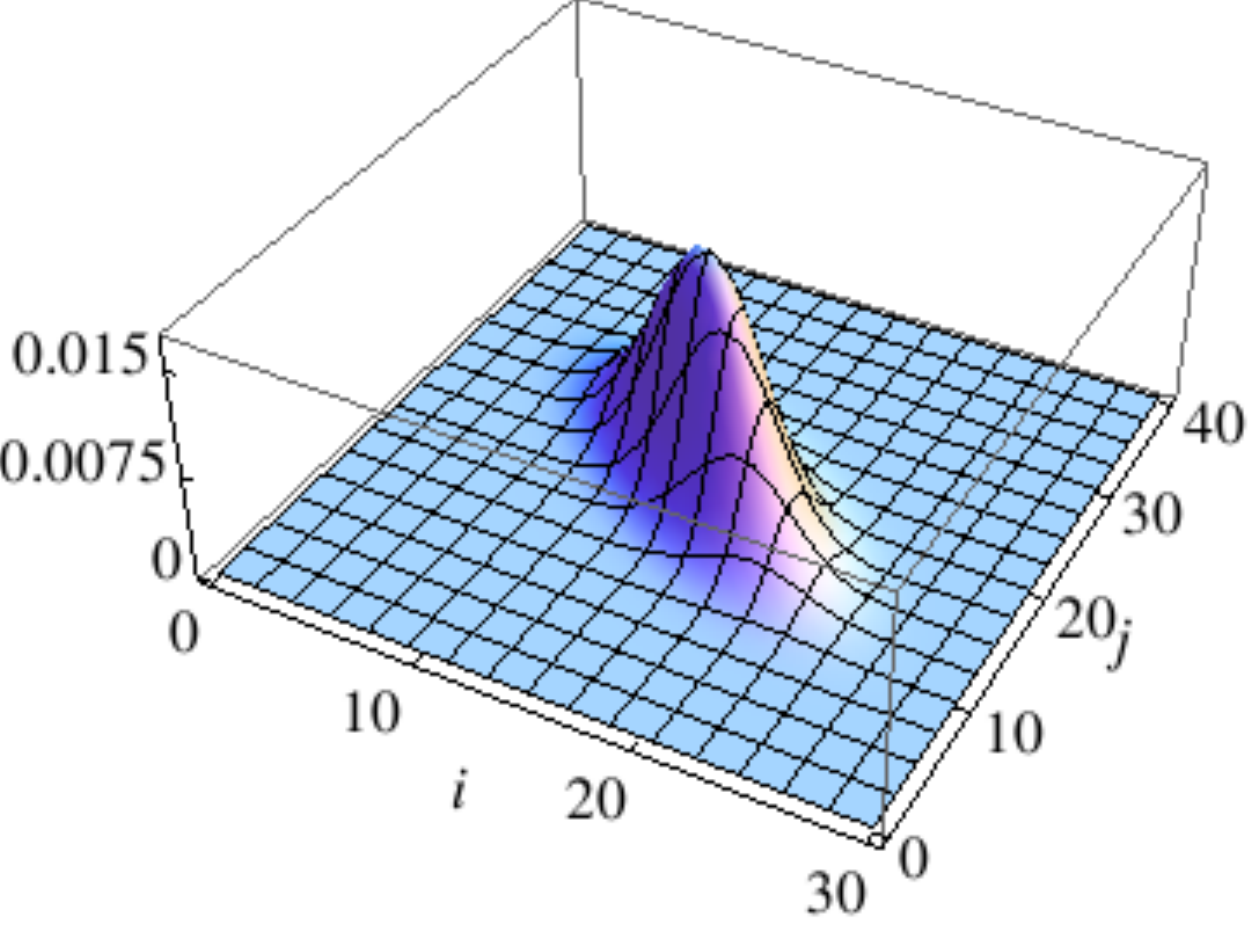}
&
\includegraphics[
clip,width=0.3\textwidth ]{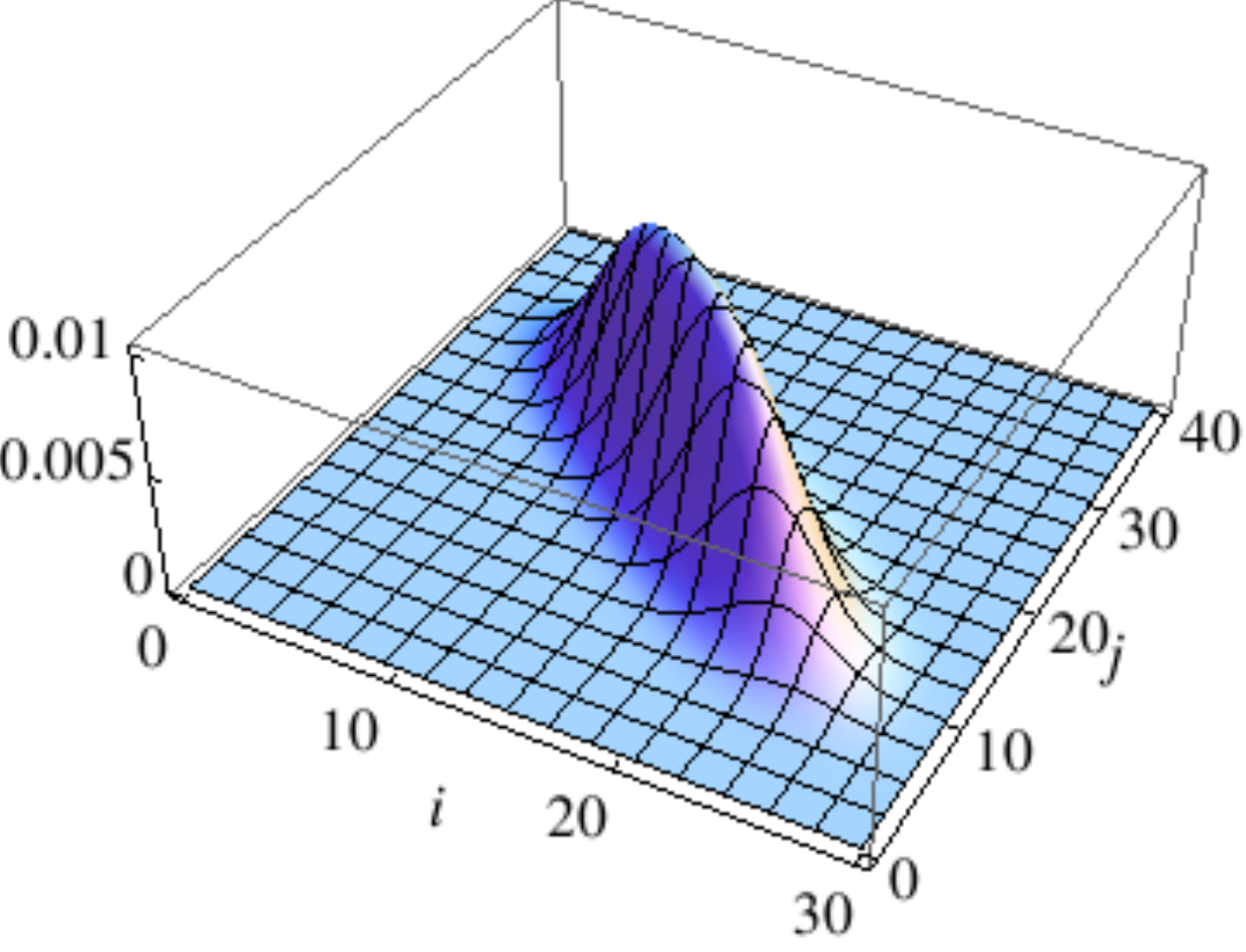}
&
\includegraphics[
clip,width=0.3\textwidth ]{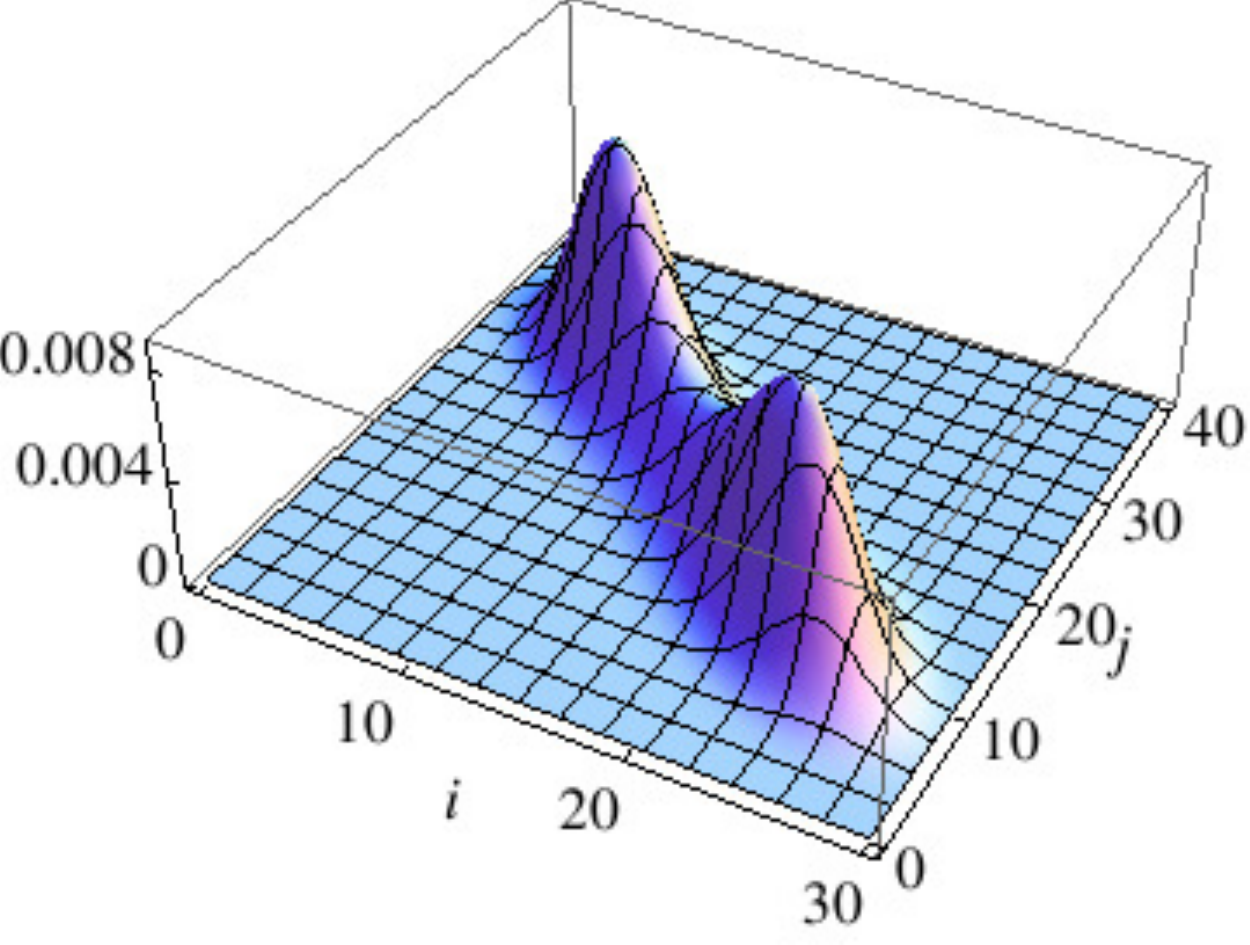}
\\
\includegraphics[
clip,width=0.3\textwidth ]{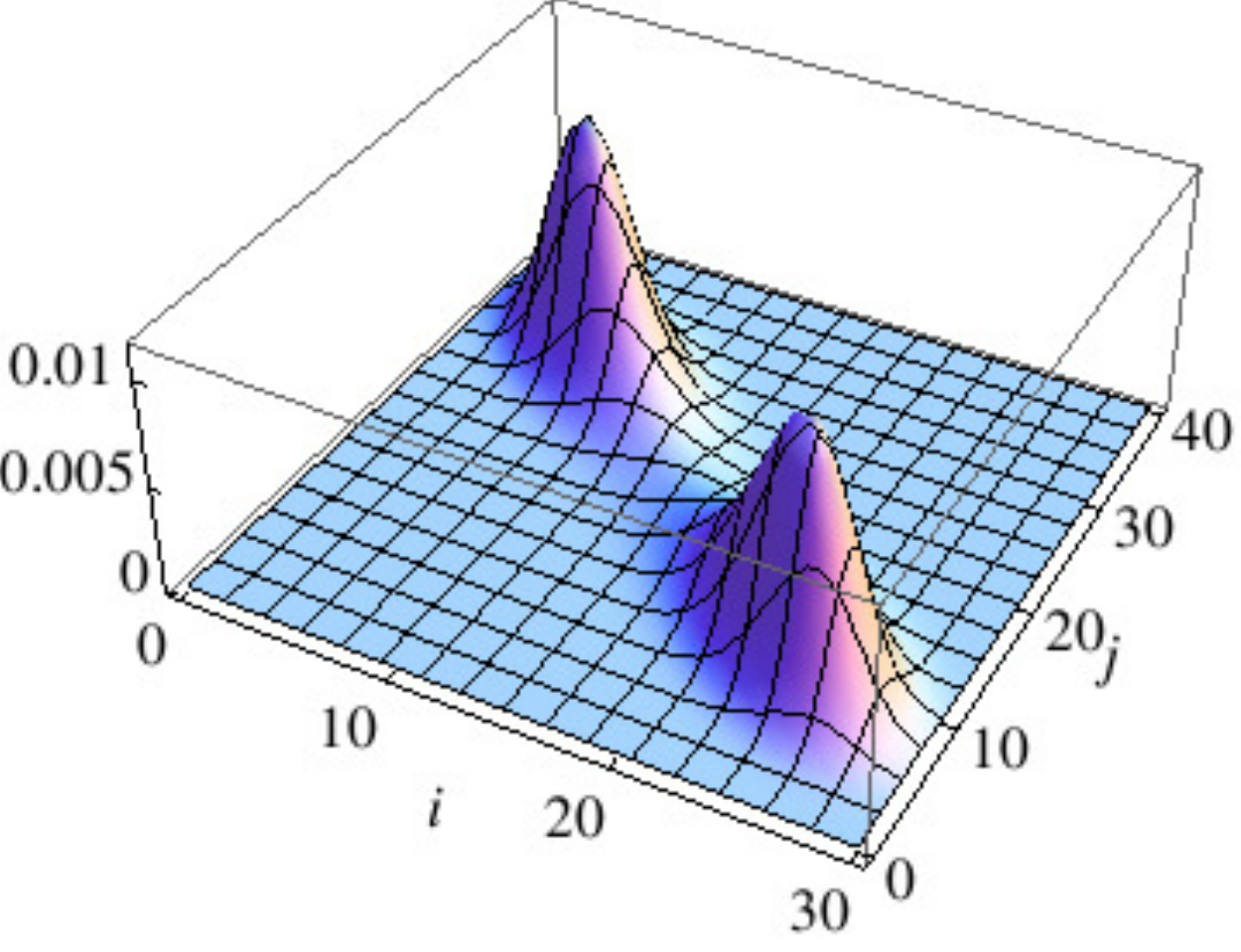}
& 
\includegraphics[
clip,width=0.3\textwidth ]{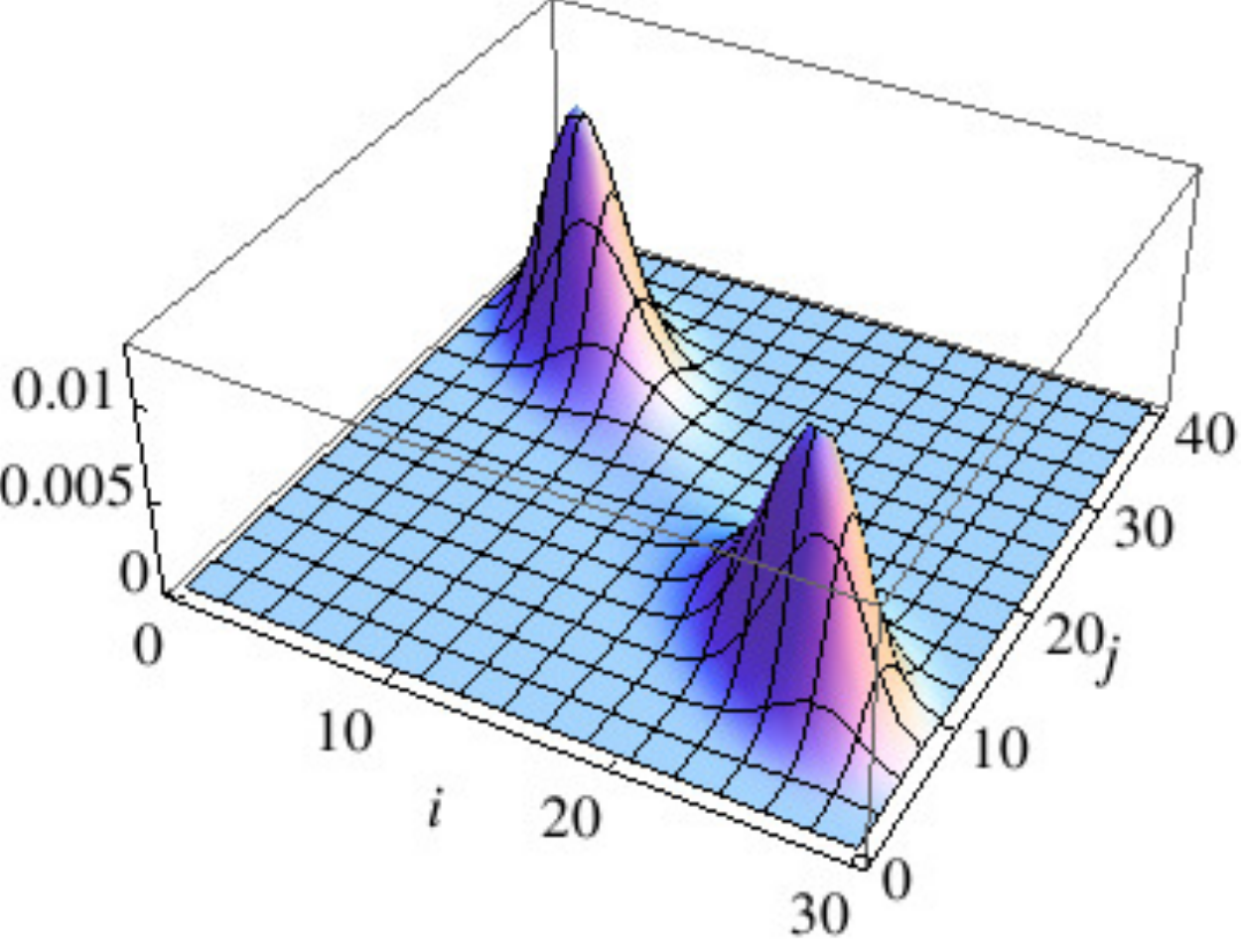}
&
\includegraphics[
clip,width=0.3\textwidth ]{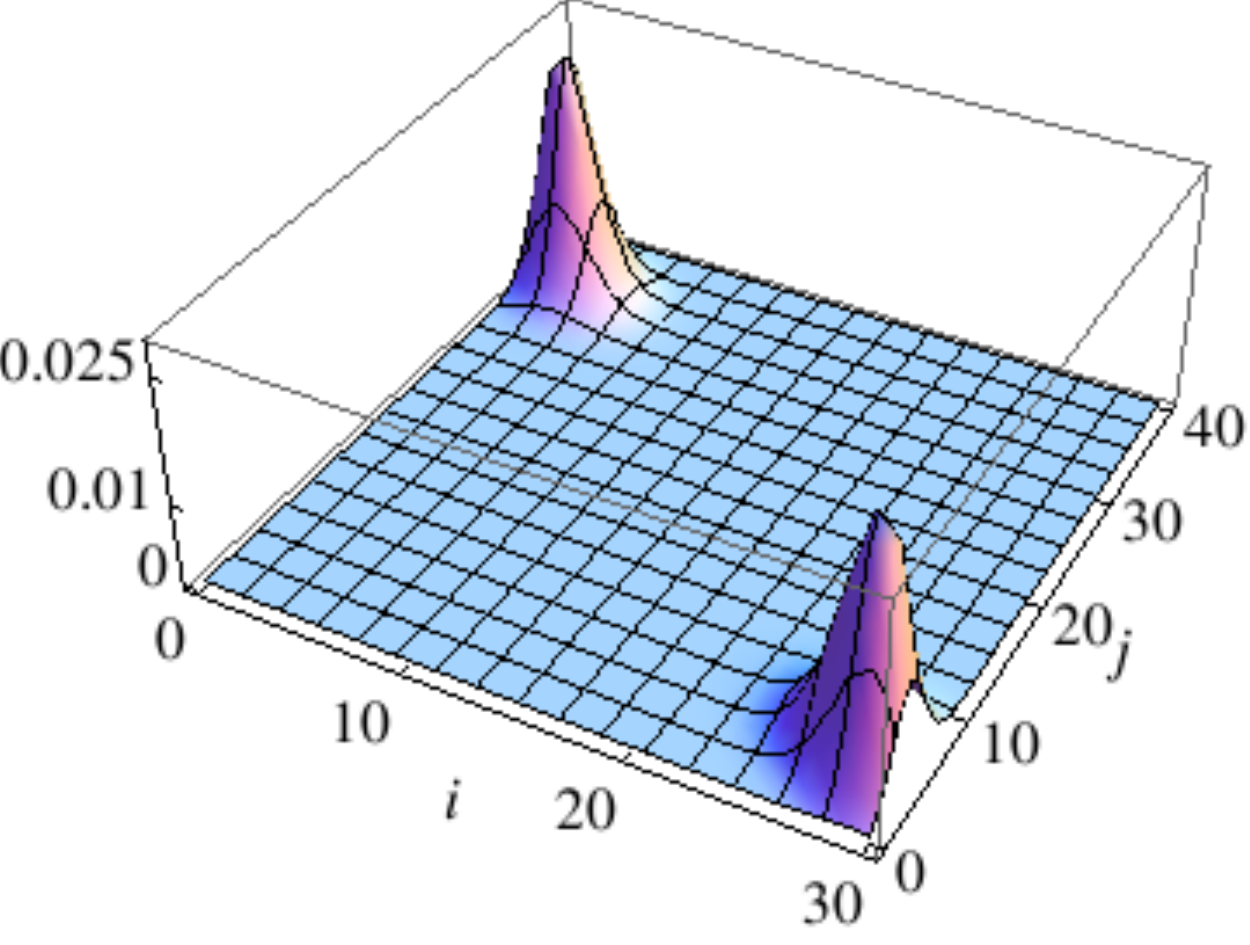}
\end{tabular}
\caption{(Color online) 
Ground-state coefficients $|C({i,j})|^2$ vs $i$ (left occupation numbers of species $a$) 
and $j$ (left occupation number of species $b$) for $U=0.1$. 
First row: 
$W=0.15$, $W=0.16$, $W=0.165$. Second row: $W=0.168$, $W=0.17$, $W=0.2$.
Number of bosons: $N_a=30$, $N_b=40$. Energies in units of $J$.}
\label{fig2n}
\end{center}
\end{figure}

\begin{figure}[h]
\begin{center}
\begin{tabular}{ccc}
\includegraphics[
clip,width=0.3\textwidth ]{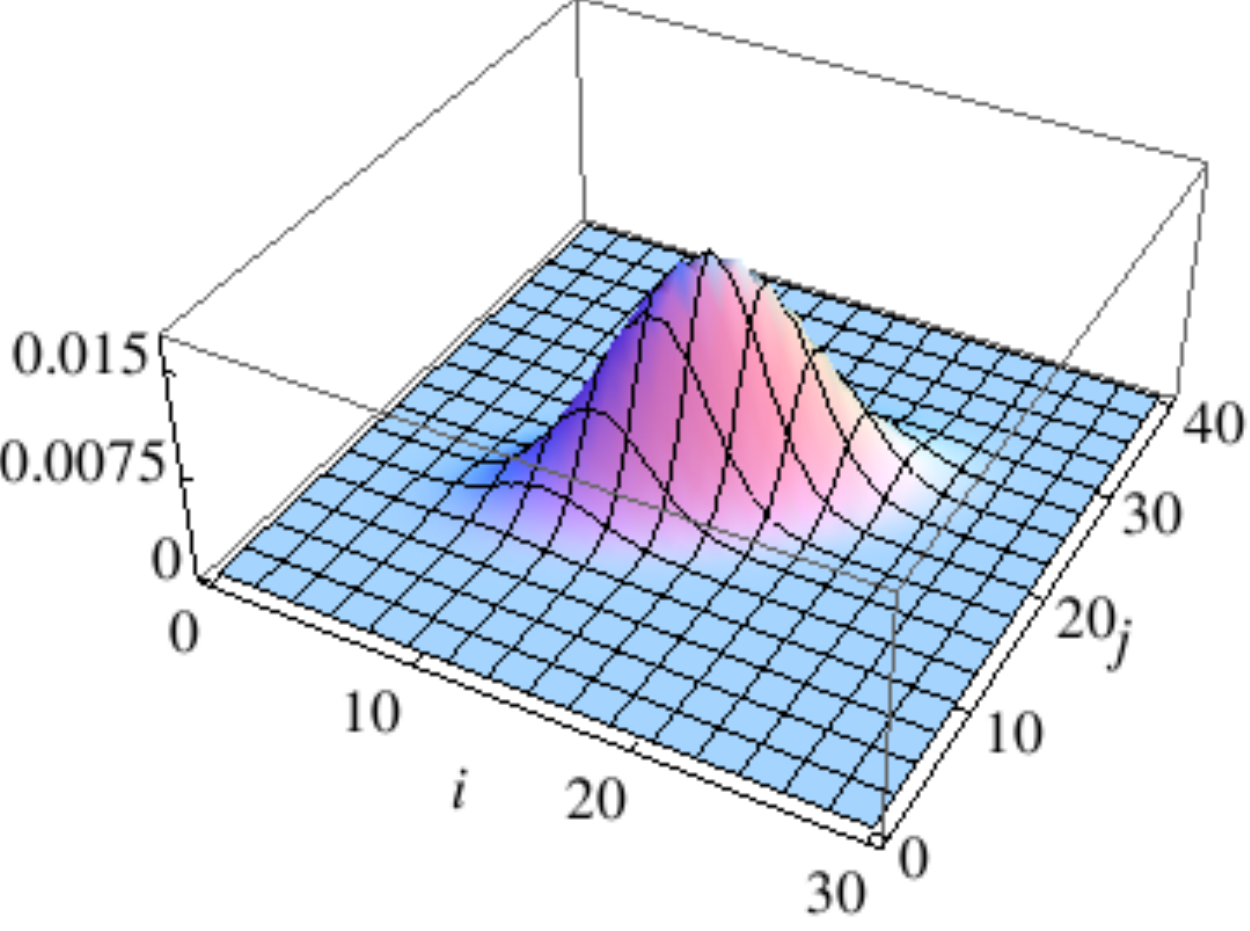}
&
\includegraphics[
clip,width=0.3\textwidth ]{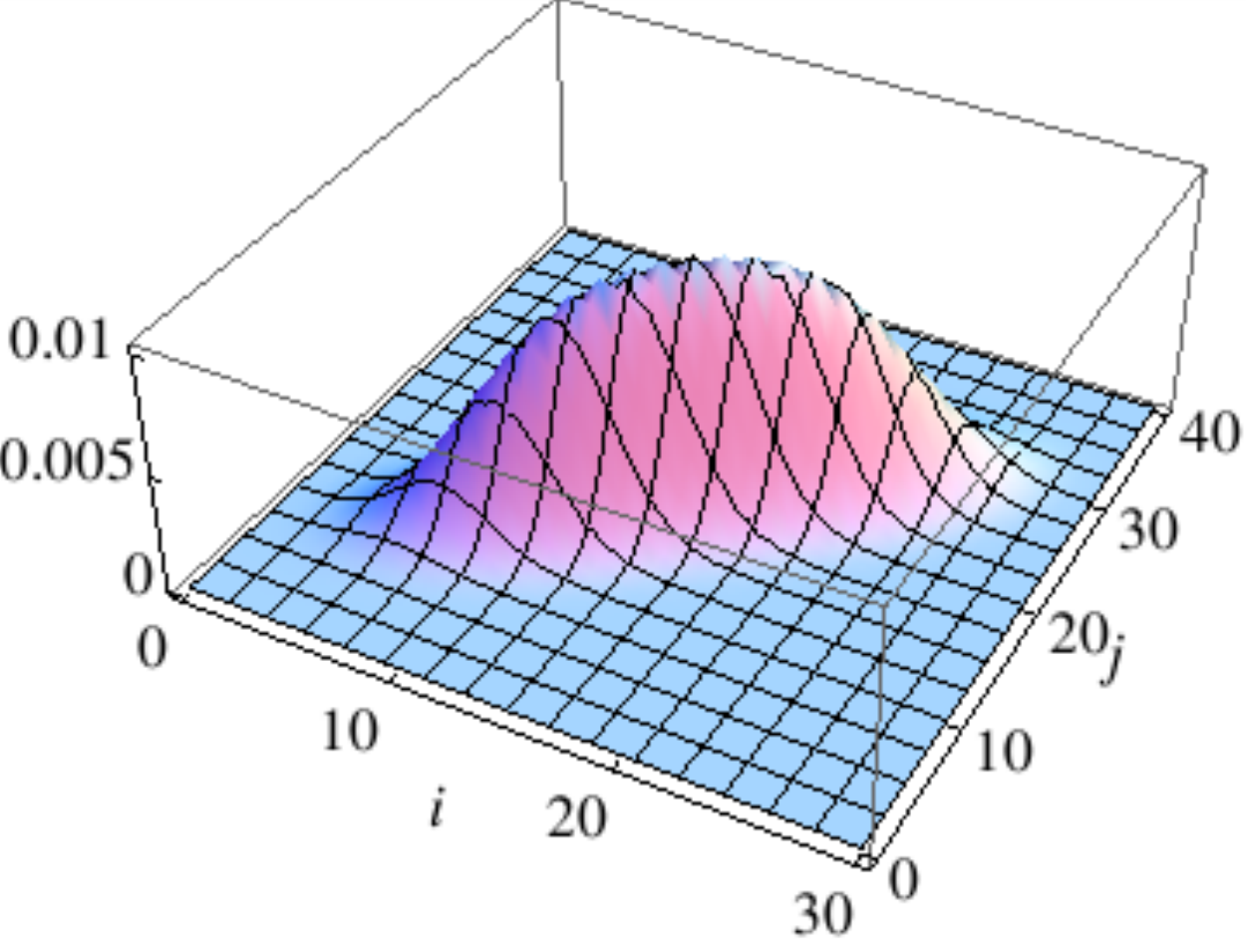}
&
\includegraphics[
clip,width=0.3\textwidth ]{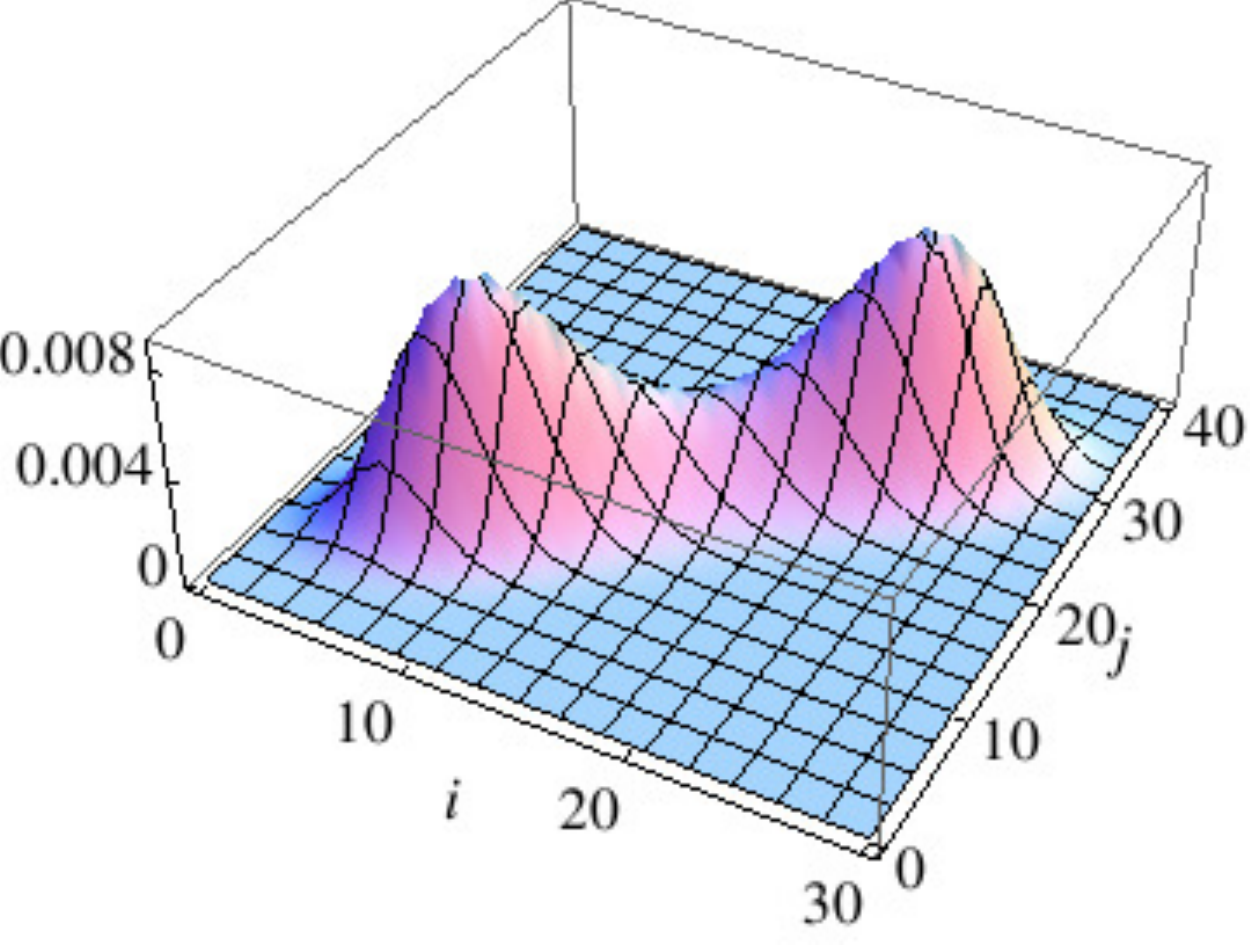}
\\
\includegraphics[
clip,width=0.3\textwidth ]{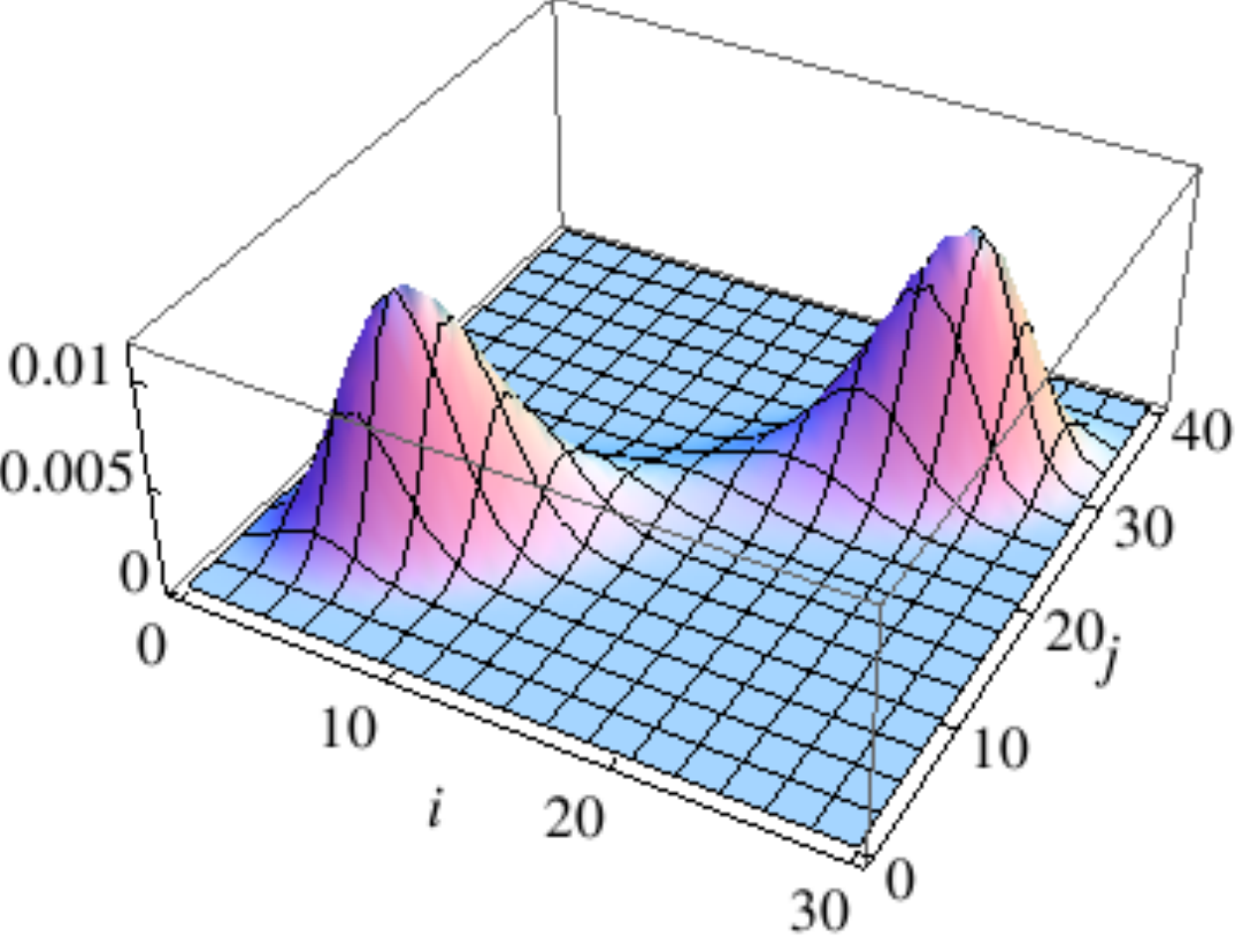}
&
\includegraphics[
clip,width=0.3\textwidth ]{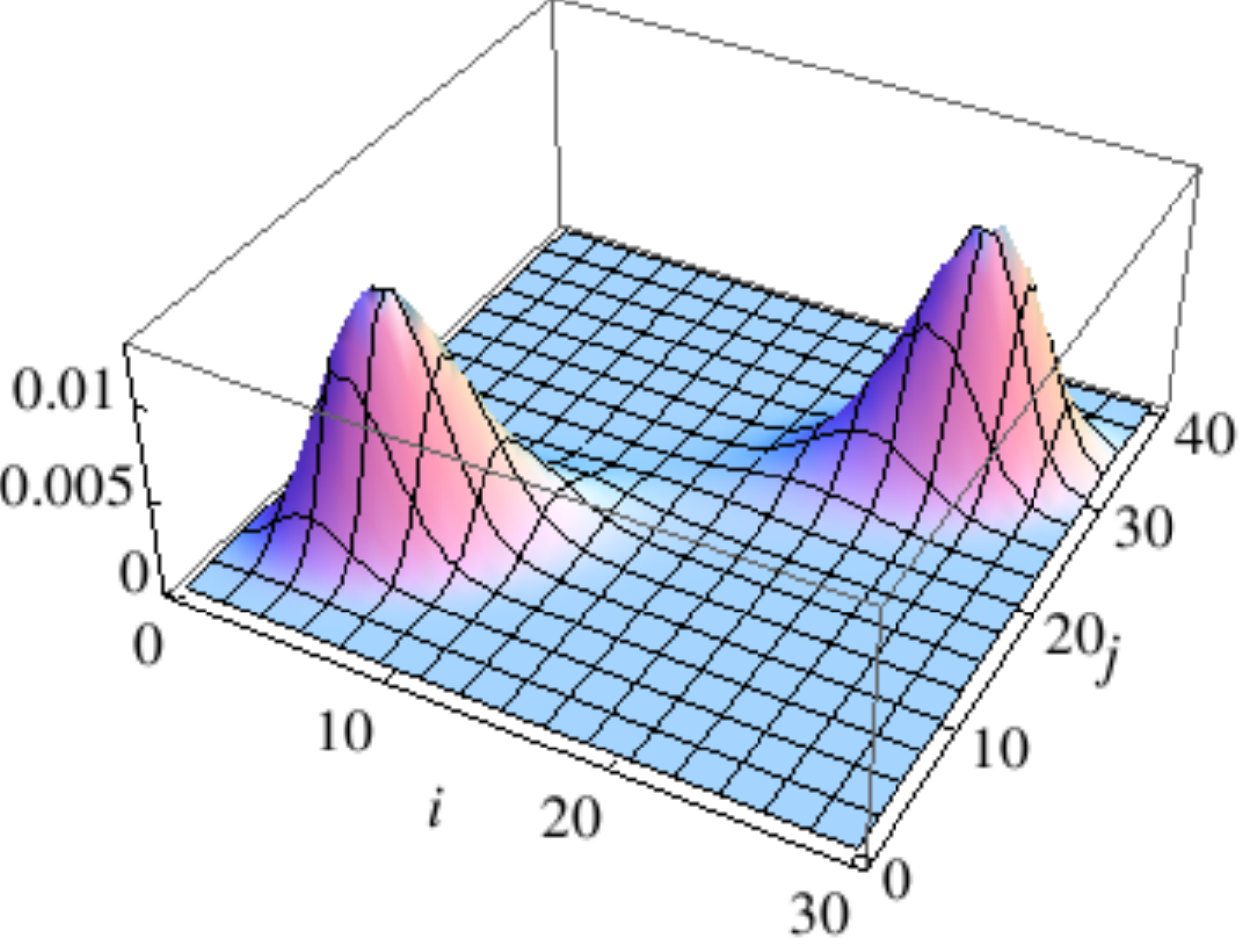}
&
\includegraphics[
clip,width=0.3\textwidth ]{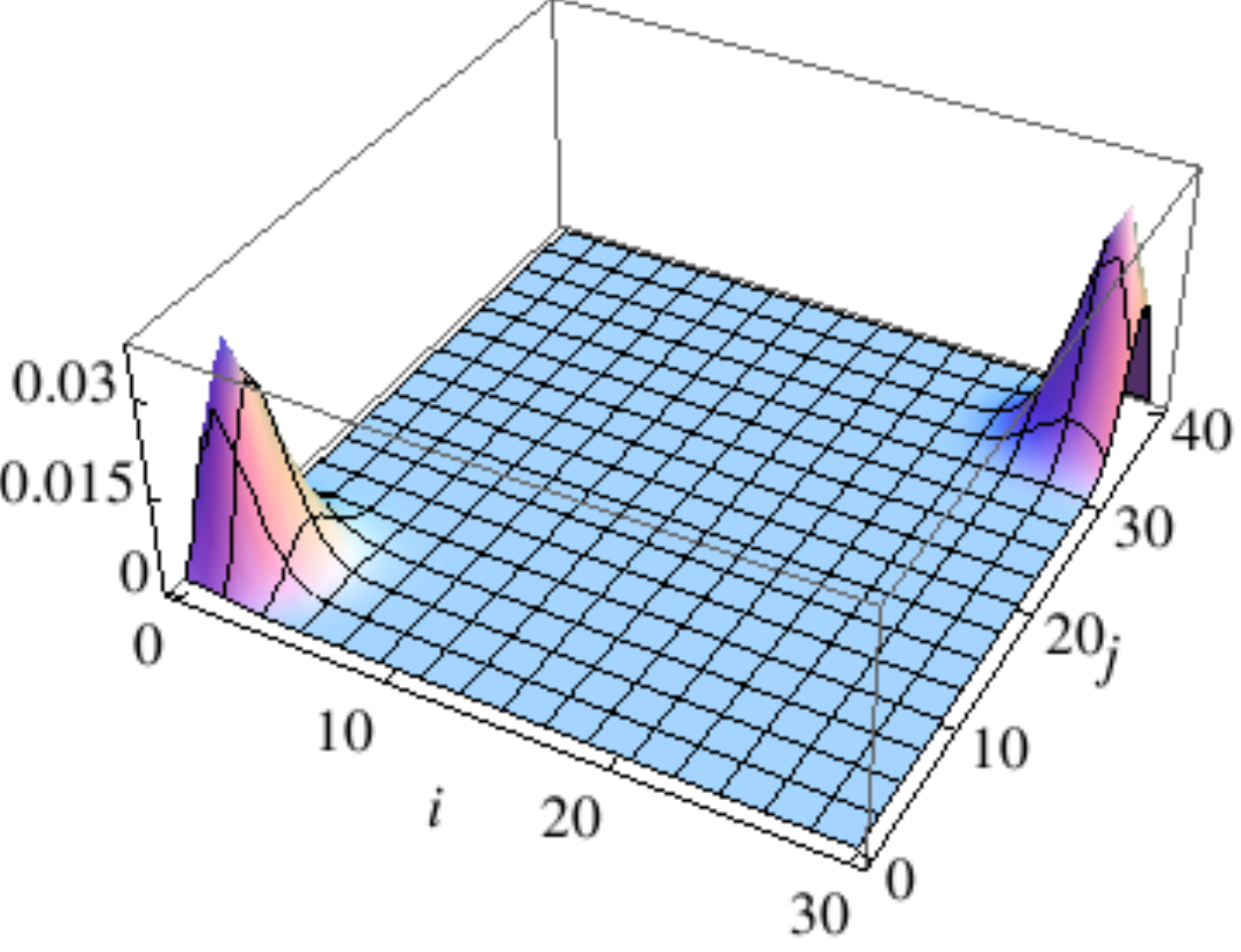}
\end{tabular}
%
\caption{(Color online) 
Ground-state coefficients $|C({i,j})|^2$ vs $i$ (left occupation number
of species $a$) and $j$ (left occupation number of species $b$) for $U=0.1$.
Top-bottom. First row: $W=-0.15$, $W=-0.16$, $W=-0.165$. Second row: $W=-0.168$,
$W=-0.17$, $W=-0.21$. Boson numbers: $N_a=30$, $N_b=40$. Energies in units of $J$.}
\label{fig3n}
\end{center}
\end{figure}

%
If one assumes that the intraspecies interactions $U_a$, $U_b$ are essentially negligible
with respect to the (absolute value of the) interspecies interaction $|W|$, 
the ground state of the TSD
Hamiltonian is essentially formed by a symmetric superposition of two states 
with macroscopically populated wells. Its form depends on the sign
of $W$. In particular, for $W>0$ (repulsive interspecies interaction), the ground state has the form
\begin{equation}
\label{catrep}
|MPW \rangle \simeq \frac{1}{\sqrt{2}}
\Bigl (
|N_a,0\rangle_L |0,N_b\rangle_R+|0,N_b\rangle_L |N_a,0\rangle_R
\Bigr )\, ,
\end{equation}
whereas for $W<0$ (attractive interspecies interaction) the ground state is
\begin{equation}
\label{catatt}
|MPW \rangle'  \simeq \frac{1}{\sqrt{2}}
\Bigl (|N_a,N_b\rangle_L |0,0\rangle_R+ |0,0\rangle_L |N_a,N_b\rangle_R
\Bigr )\; .
\end{equation}
In Figs. \ref{fig1n}, \ref{fig2n} and \ref{fig3n},
we show the ground state of Hamiltonian (\ref{bhtwo}) when $W$ is varied.
The left panel of Fig. \ref{fig1n}
represents the ground state when boson-boson interactions are zero, namely, $U=0=W$.
This situation corresponds to the atomic coherent state (\ref{coherentdouble}).
As shown in the right panel of the same figure for which $W=0$ and $U= 0.1 J$,
the distribution $|C({i,j})|^2$ becomes narrower as soon as
the on-site repulsive interaction is switched on.
%
%
By keeping fixed $U=0.1 J$, a finite and repulsive interaction $W$ balances this effect
and causes the broadening of distribution $|C({i,j})|^2$ which becomes more and more pronounced
when $W$ is increased, as the left panel of the first row in Fig. \ref{fig2n} shows.
%
By further increasing the interspecies repulsion, the ground-state structure exhibits
a transition to a configuration characterized by the coexistence of a delocalized state
(well represented by state (\ref{coherentdouble})) and the two localized states
$|N_a, 0\rangle_L |0,N_b\rangle_R$ and $| 0,N_b\rangle_L |N_a,0\rangle_R$.
Such an effect (already observed for bosons in a three-well potential with ring geometry
\cite{vittorio1}) is well visible in the first row (right plot) of Fig. \ref{fig2n}.
%
%
This coexistence (where the ground state losses its coherence) is destroyed by
a sufficiently-strong interspecies repulsion. In this case, the ground state
becomes the symmetric superposition of (macroscopic) localized states described
by state (\ref{catrep}) and well illustrated in the second row (right plot) of Fig. \ref{fig2n}.
In correspondence to the same value of the onsite interaction $U$ but with attractive interspecies
interaction ($W<0$, see Fig. \ref{fig3n}), distribution of $|C({i,j})|^2$ displays the same
changes characterizing the repulsive case when $W$ is increased.
Fig. \ref{fig3n} clearly shows that sufficiently strong interspecies attractions give rise to 
a superposition having the form of state (\ref{catatt}) describing localized mixed species.

\section{Ground-state correlation properties}

\begin{figure}[h]
\begin{center}
\includegraphics[
clip,width=0.7\textwidth ]{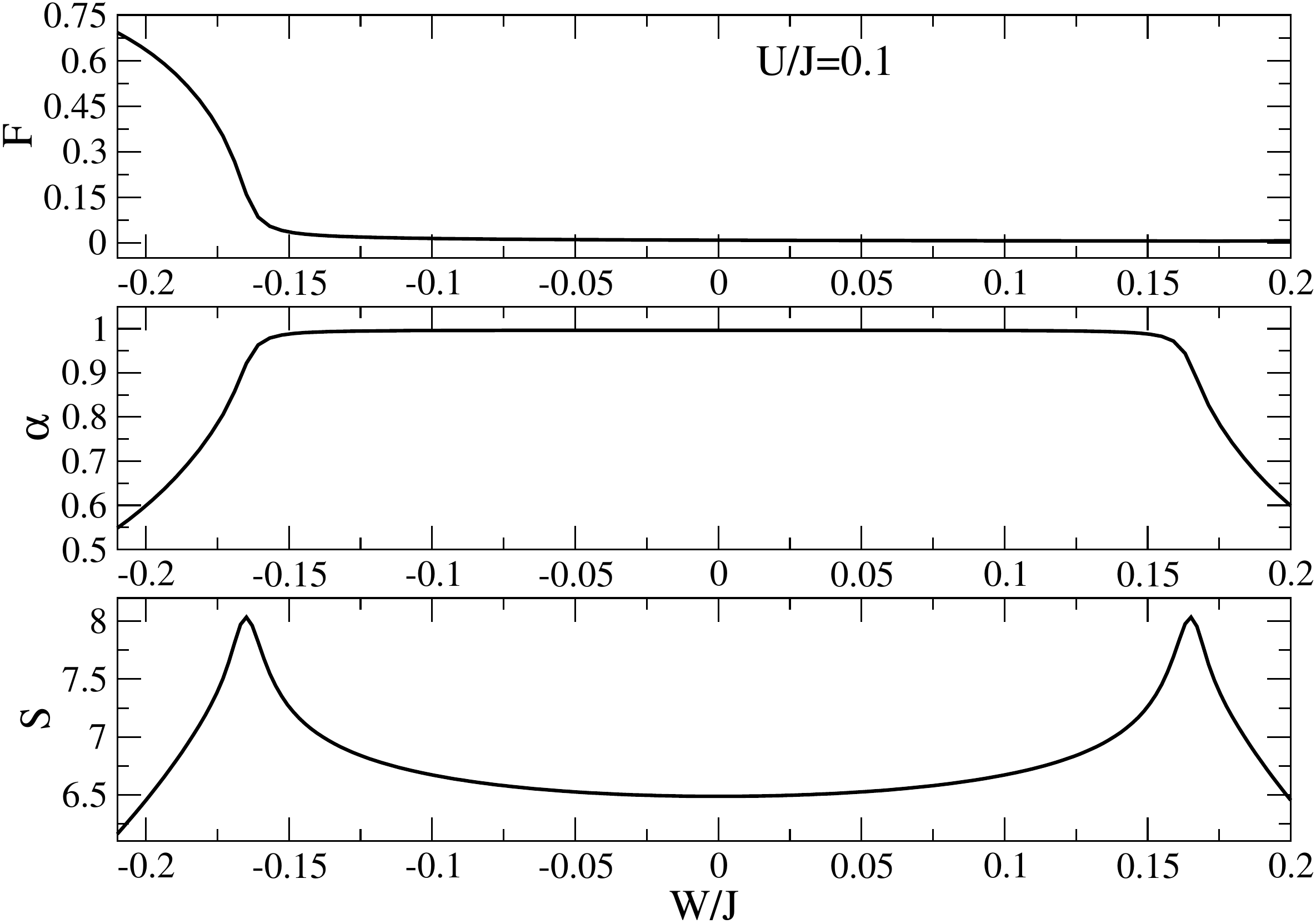}
\caption{(Color online) 
Top panel: normalized Fisher information $F=\mathcal{F}/N^2$ (with $\mathcal{F}$
given by Eq. (\ref{fisher})) vs $W/J$. Middle panel: coherence visibility $\alpha$
(see Eq. (\ref{coherence})) vs $W/J$. Bottom panel: entanglement entropy (\ref{ee})
vs $W/J$. In all three panels: $U/J=0.1$. Number of bosons:
$N_a=30$, $N_b=40$. All the quantities are dimensionless.}
\label{fig4n}
\end{center}
\end{figure}

We study the correlation properties of the ground state by following the same path
followed in Ref. \cite{mazzpar}. We thus use parameters that are appropriate extensions
to the case under investigation of the Fisher information \cite{braunstein,pezze}, 
the coherence visibility \cite{stringari,anglin,anna} and the entanglement entropy \cite{bwae}.
%
The Fisher information is defined by
\begin{equation}
\label{fisher}
\mathcal{F}=\langle (\hat{N}_L-\hat{N}_R)^2 \rangle
\; ,
\end{equation}
where the averages (here and in the following) are intended calculated
with respect to the ground state, and
$\hat{N}_{v}= {\hat{a}}^\dagger_{v} \hat{a}_{v}
+{\hat{b}}^\dagger_{v} \hat{b}_{v}$ ($v=L,R$)
is the boson number in the well $v$.
The right-hand side of Eq. (\ref{fisher}) provides a measure of the fluctuations of the interwell
boson population imbalance due to the collective particle transfer caused by the tunneling 
of bosons across the central barrier. Note that the Fisher information is related to
those entanglement properties descending from the quantum indistinguishability of the particles, 
a property known as multiparticle entanglement \cite{pezze}.
The quantity $\mathcal F$ is equal to $(N_a+N_b)^2=N^2$ when the ground state is
state (\ref{catatt}) and to $(N_a-N_b)^2 \equiv (\Delta N)^2$ in correspondence to
state (\ref{catrep}).
Thus we use the indicator $\mathcal{F}$  to have a feedback for the occurrence of symmetric superpositions
of two states with macroscopically populated sites.
We analyze the coherence properties of the ground state due to the single-particle hopping
from the left well to the right one and back. To do this we use the following weigthed 
coherence visibility
%
%
\begin{equation}
\label{coherence}
\alpha=\frac{N_a\,\alpha_a+N_b\,\alpha_b}{N_a+N_b}
\end{equation}
$$
\alpha_a= 2\frac{\langle \hat{a}^{\dagger}_{L} \hat{a}_{R}\rangle}{N_a}
\; ,\quad
\alpha_b= 2\frac{\langle \hat{b}^{\dagger}_{L} \hat{b}_{R}\rangle}{N_b}
\;,
$$
where $\alpha_c$ is the visibility (of the intereference fringes in the ground-state momentum 
distribution) of the compoment $c$ \cite{stringari,anglin,anna}, \cite{mazzpar}.
In the spirit of the single-particle tunneling, one can see, for example, from the first 
equation of the second row of Eq. (8), that the operator $\hat{a}^{\dagger}_{L}\hat{a}_{R}$ 
destroys a single boson in the right well and creates it in the right well.
The coherence (8) is equal to one when the ground state is the state (4). Thus we use 
the indicator $\alpha$ to signal the emergence of the atomic coherent state.
%
%
We study to what extent the ground state $|\psi\rangle$ (described by the density matrix 
$\hat{\rho}=|\psi\rangle \langle \psi|$) is affected by the genuine quantum correlations 
(entanglement) between the left well and the right one.
%
%
We carry out this analysis from the bi-partition perspective with the two wells playing the role of the two partitions. Since the system is in a pure state, an excellent measure of the entanglement between the two wells is the entanglement entropy $S$. 
%
%
%
This quantity is obtained by first tracing out the degrees of freedom of the (right) left well from the density matrix $\hat{\rho}$ 
so as to get the reduced density matrix $\hat{\rho}_{L(R)}$, and, then, calculating the von Neumann entropy of the afore reduced
density matrix by evaluating the trace of the matrix $-\hat{\rho}_{L(R)}\log_{2}\hat{\rho}_{L(R)}$. 
The entanglement entropy for our systems is thus described by
\begin{equation}
\label{ee}
S=-\hat{\rho}_{L(R)}\log_{2}\hat{\rho}_{L(R)}=
-\sum_{i=0}^{N_a}\sum_{j=0}^{N_b}|C_{i,j}|^2 \log_{2}|C_{i,j}|^2
\; ,
\end{equation}
where $C_{i,j}$ are the coefficients of the ground-state expansion in terms of Fock states [see Eq. (3)] that we determine numerically.
The eventual maxima of $S$ as a function of the boson-boson interaction will signal the ground 
states featuring the highest achievable degrees of left-right pure quantum correlations. As for the 
single-component case [44], these interactions might sign the onset of the self-trapping regime in 
the dynamics of the corresponding two-species bosonic Josephson junction.
In Fig. \ref{fig4n} we have plotted the normalized Fisher information $F= \mathcal{F}/N^2$
(with $\mathcal{F}$ given by Eq. (\ref{fisher})), the visibility $\alpha$
(see Eq. (\ref{coherence})), and entropy (\ref{ee})
as functions of interspecies interaction $W$ by keeping $U$ fixed. From this figure,
we observe that $F$ attains its maximum when the ground state has the shape of the
superposition (\ref{catatt}), see the right panel in the second row of 
Fig. \ref{fig3n}, where $W/J=-0.21$.

The visibility $\alpha$ exhibits an interesting behavior heralded by the analysis
displayed in Figs. \ref{fig2n} and \ref{fig3n}. This shows that the ground state of
Hamiltonian (\ref{bhtwo}) is characterized by
a high degree of coherence over a finite range of values of interaction $W$.
The boundaries of this region are the two values of $W$ signing the on-set of the coexistence
of the delocalized state and the symmetric superposition of two localized states.
In correspondence to such values of $W$, $S$ features a double-peak structure characterized 
by two maxima, as shown by the bottom panel of Fig. \ref{fig4n}. The von Neumann entropy thus
confirms its effectiveness by measuring the increased degree of entanglement characterizing
states (\ref{catrep}) and (\ref{catatt}). 

\section{Low-energy spectrum} 
\label{wes}

The analysis of section \ref{qa} about the different regimes characterizing the ground state
provides useful information for studying the spectrum of the model Hamiltonian through the 
Bogoliubov approach.
For example, in the two strong-interaction regimes ($|W|>U_c$) relevant to $W > 0$ (repulsive
interaction) and $W <0$ (attractive interaction), the macroscopic localization effects
illustrated in Figs. \ref{fig2n} and  \ref{fig3n}, respectively, show how
for $W>0$ one observes
an almost complete separation of the two species strongly localized
in different sites, whereas for $W <0$ one has an almost complete merging
of the two species localized at the same site. This allows one to identify
weakly-occupied modes and thus to implement
the Bogoliubov scheme.

A more complicated situation is found in the regimes $|W|< U_c$ (again including
the cases $W < 0$ and $W>0$), where the two bosonic species are
completely delocalized and the boson populations essentially exhibit the same size in the
two wells. In this second case, the correct approach for diagonalizing the Hamiltonian
is found by looking for macroscopically-occupied modes within the momentum-mode picture.

\subsection{Low-energy states with delocalized populations and spectral collapse}
\label{51}

First we discuss the repulsive case 
of the two cases with $|W|<U_c$ involving
delocalized populations and species mixing. Since the two species are almost equally
distributed in the two wells
a simple way to recognize weakly-occupied modes \cite{vittorio2013}
is to introduce momentum-like operators
$\hat{A} = { ( \hat{a}_{L} + \hat{a}_{R}) }/{\sqrt 2}$,
$\hat{\alpha} = { (\hat{a}_{L} - \hat{a}_{R}) }/{\sqrt 2}$,
$\hat{B} = {(\hat{b}_{L} + \hat{b}_{R} )}/{\sqrt 2}$
and
$\hat{\beta}  = {( \hat{b}_{L} - \hat{b}_{R} )}{\sqrt 2}$.
%
%
Note that the boson-number conservation is now given by
$$
\left
\{
\begin{array}{ll}
{N}_a = N_{aR}+N_{aL} = N_A +N_\alpha  \, ,
\\
{\-}
\\
{N}_b = N_{bR}+N_{bL} = N_B +N_\beta \, .
\end{array}
\right .
$$
In the current regime ($|W|<U_c$), modes $\hat{A}$, $\hat{B}$ are macroscopically occupied,
($N_A \gg N_\alpha $, $N_B \gg N_\beta $) and can be regarded as complex parameters within
the Bogoliubov approximation. The application of this scheme (see \ref{uno}) provides
the diagonal Hamiltonian
\begin{equation}
\hat{H}_D = K -2J + \sqrt {R_\alpha } \;
(2 {\hat \alpha}^+ {\hat \alpha}+1)+ \sqrt {R_\beta } \; (2{\hat \beta}^+{\hat \beta}  +1)
\; ,
\label{bogD}
\end{equation}
where $K = \sum_{c= a,b} {U_c}({N}^2_c -{N}_c)/{4}$$+{W} {N}_a {N}_b/{2} -JN$, and
$J_a = J_b = J$ has been assumed. The symbols ${R_\alpha }$ and ${R_\beta}$ are defined by
$$
R_{\alpha,\beta} = J [J + (u \mp D)/4 ]\; ,\quad D =\sqrt{ \Delta^2 +4W^2N_aN_b }\; , 
$$
with signs $-$ ($+$) corresponding to $R_{\alpha}$ ($R_{\beta}$),
$u = U_aN_a +U_bN_b$, and $\Delta = U_a N_a -U_bN_b$.
Note that, we have tacitly assumed
$U_bN_b >U_aN_a$. In the opposite case $U_bN_b <U_aN_a$
the definitions of $R_\alpha$, $R_\beta$ are simply exchanged.
%
%
The eigenvalues read
\begin{equation}
\label{eigenvalueswsu}
E(N_\alpha, N_\beta )\! = \!
K -2J  + \sqrt {R_\alpha}(2{N}_\alpha+1) +  \sqrt {R_\beta}(2{N}_\beta+1)
\end{equation}
and the corresponding energy eigenstates
\begin{equation}
\label{eigenvectorswsu}
|E(N_\alpha , N_\beta)\rangle = |N_A \rangle | N_B \rangle
\hat{U}^\dagger \hat{S}_\alpha \hat{S}_\beta |N_\alpha \rangle | N_\beta \rangle
\end{equation}
are obtained by mixing the squeezed Fock states
$\hat{S}_\alpha |N_\alpha \rangle$ and $\hat{S}_\beta | N_\beta \rangle$ 
($\hat{S}_\alpha$ and $\hat{S}_\beta$ are, in fact, squeezing operators) through the
SU(2)-group rotation $\hat{U}$. These states, $\hat{U}$ and the SU(1,1)-group 
transformations $\hat{S}_\alpha$ and $\hat{S}_\beta$ are defined in \ref{uno}.

As far as $4J +u > D$,
the quantity $R_\alpha$ is positive
and the contributions to the spectrum relevant to $N_\alpha$ and $N_\beta$ are both discrete.
A dramatic change in the energy spectrum emerges when $R_\alpha \to 0$. By rewriting the condition
$4J +u > D$ in the form $U_aU_b + \Delta_J > W^2$, and observing that $\Delta_J= 8J (J+u)/{N_aN_b} \simeq 0$ for
bosonic populations large enough, this effect takes place when
\begin{equation}
U_aU_b -W^2 \to 0^+\; .
\label{lim}
\end{equation}
The latter formula reproduces for $U_a=U_b= U$ (twin species) the well-known mixing condition 
$U =W$ characterizing bosonic mixtures and derived heuristically in Ref. \cite{lingua}
for large-size lattices. 
For $W^2$ approaching  $U_aU_b$ from below, the interlevel separation for the $\alpha$-mode spectrum
tends to vanish thus determining, for $W^2 = U_aU_b $, the transition to a continuous spectrum. 
Hamiltonian (\ref{albe}) (emerging from the Bogoliubov approach and leading to
the diagonal form (\ref{bogD})) thus predicts
a {\it spectral collapse} of the energy levels for $W^2 \to U_aU_b$.
This mechanism is discussed in detail in \cite{vittorio2013} and \cite{ijqi}.
The resulting macroscopic effect (also observed in Ref. (\cite{felicetti})) can be 
interpreted as the hallmark of the transition from
the mixed-species regime (illustrated in the first row, left panel of Fig. \ref{fig2n})
to the demixed regime involving, for $W>0$, the spatial separation of the two species
(illustrated in the second row, right panel of Fig. \ref{fig2n}).
%
%
\begin{figure}[h]
\begin{center}
\begin{tabular}{cc}
\includegraphics[
clip,width=0.5\textwidth ]{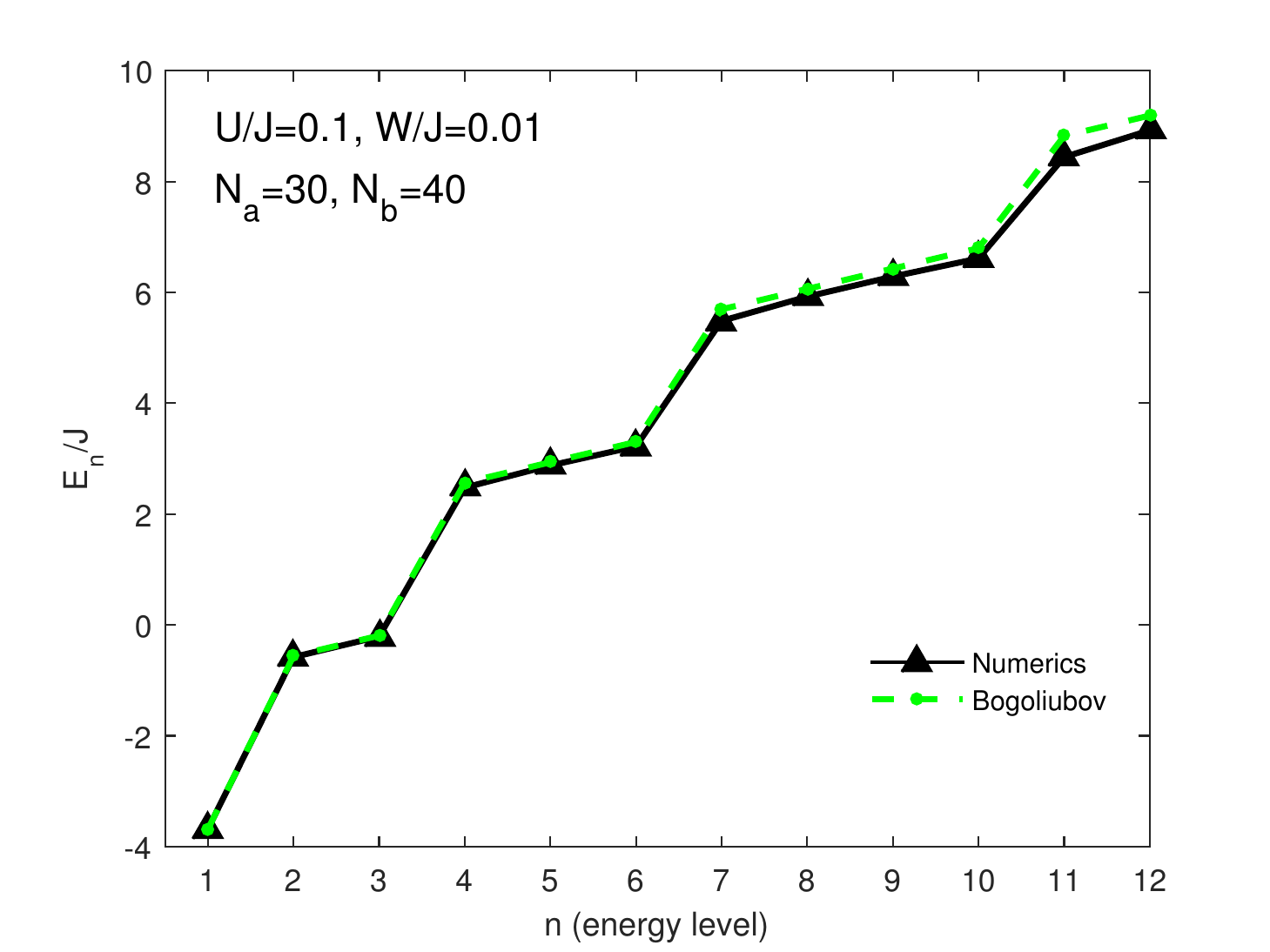}
&
\includegraphics[
clip,width=0.5\textwidth ]{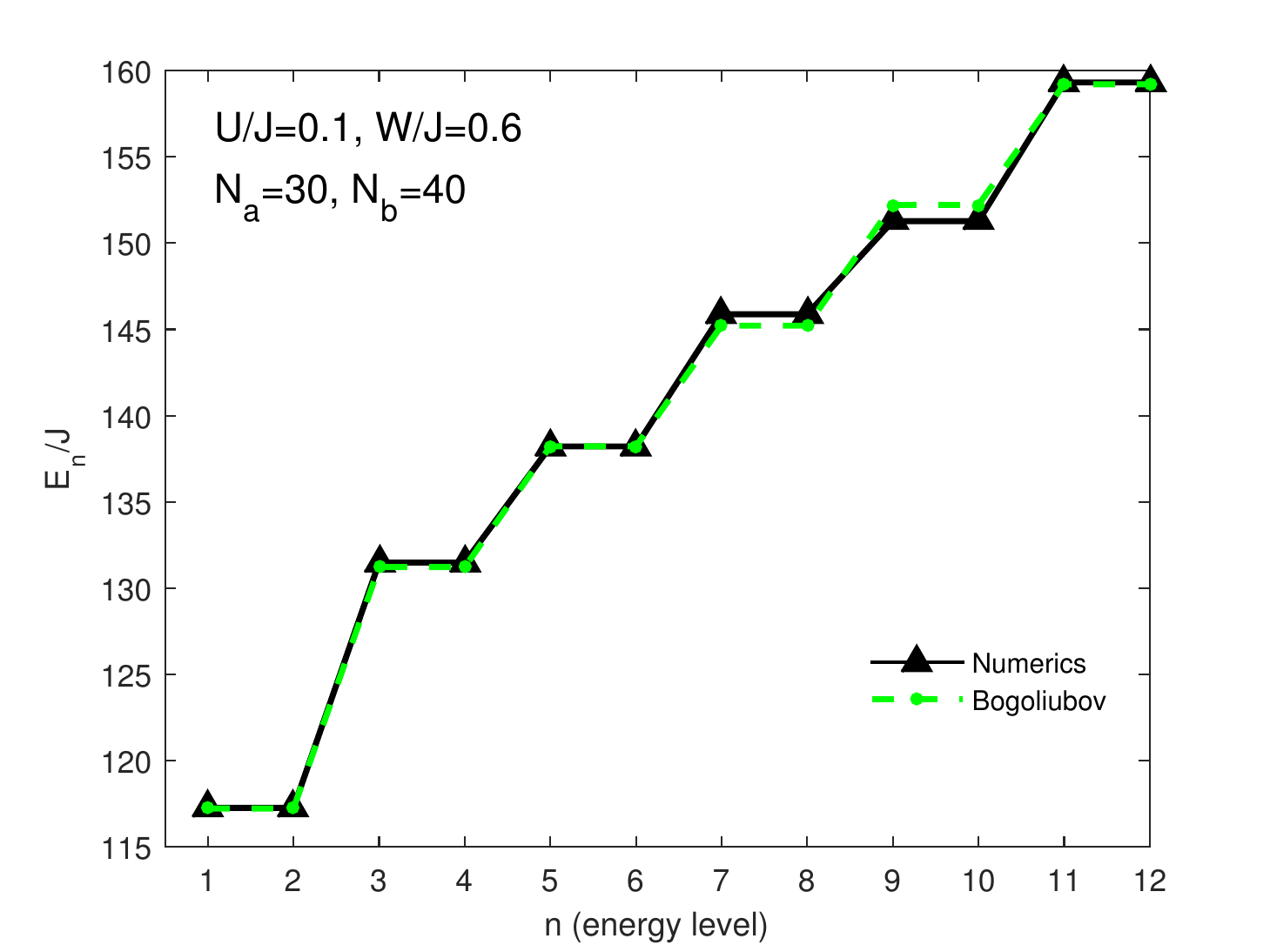}
\end{tabular}
\caption{(Color online) 
Case $W>0$ with $U_a=U_b \equiv U$. Left panel: $W<U$. Black triangles: eigenvalues obtained numerically. 
Green dots: Eigenvalues obtained by Eq. (\ref{eigenvalueswsu}),
$(N_\alpha, N_\beta)=(0,0),(1,0),(0,1),(1,1),(2,1),...$ correspond to 
the energy level $n=1,2,3,4,5,...$ ($n=1$ denotes the ground state). 
Right panel: $W>U$. Black triangles: eigenvalues obtained numerically. Green dots: Eigenvalues obtained 
by Eq. (\ref{eigenvalueswgu1}), 
$(N_{aR}, N_{bL})=(0,0),(0,1),(1,0),(1,1),(1,2),...$ 
correspond to the energy level $n=1,2,3,4,5,...$.
All the quantities are dimensionless. Lines are for eye-guide.}
\label{fig6x}
\end{center}
\end{figure}
\begin{figure}[h]
\begin{center}
\begin{tabular}{cc}
\includegraphics[
clip,width=0.5\textwidth ]{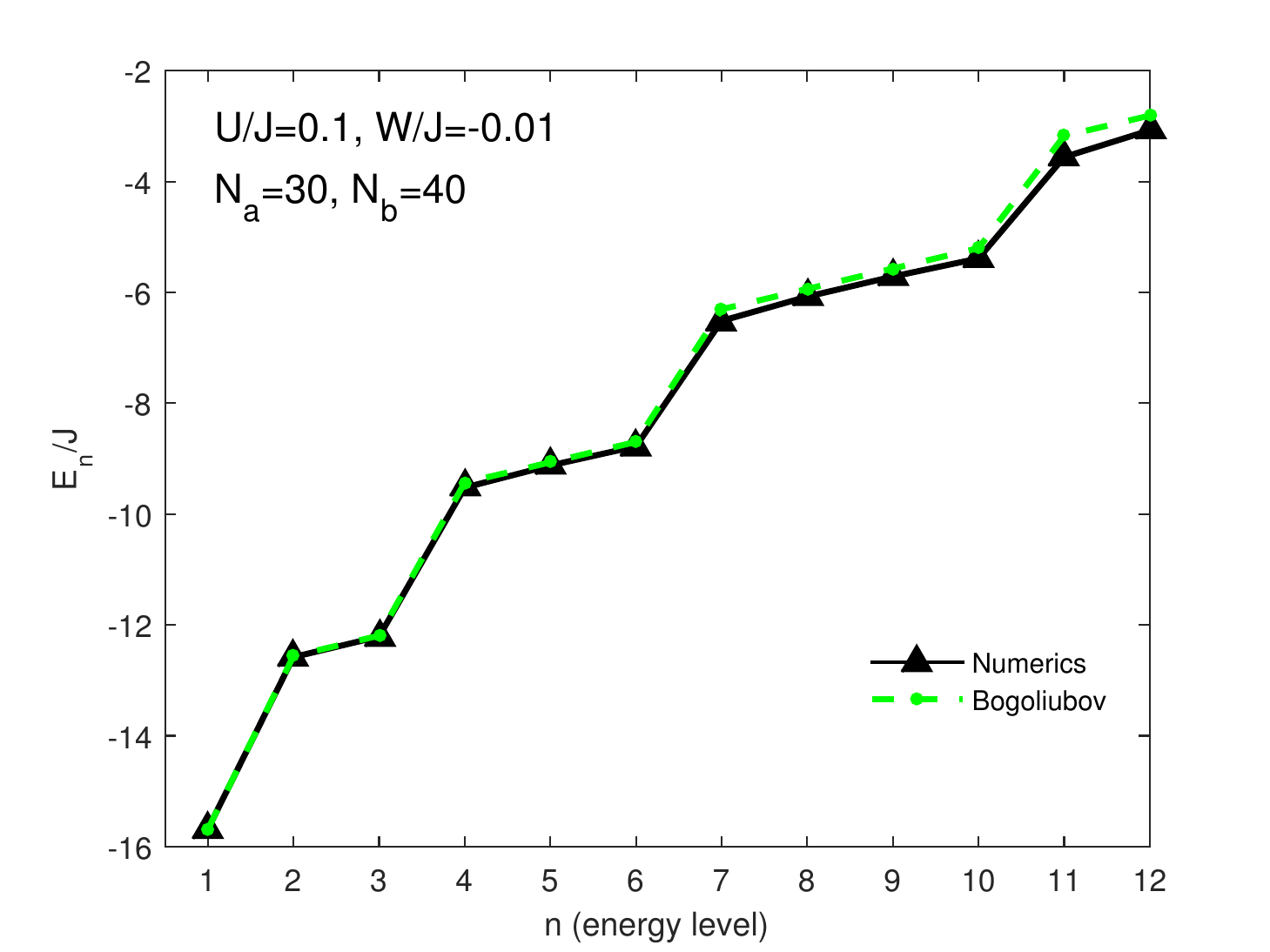}
&
\includegraphics[
clip,width=0.5\textwidth ]{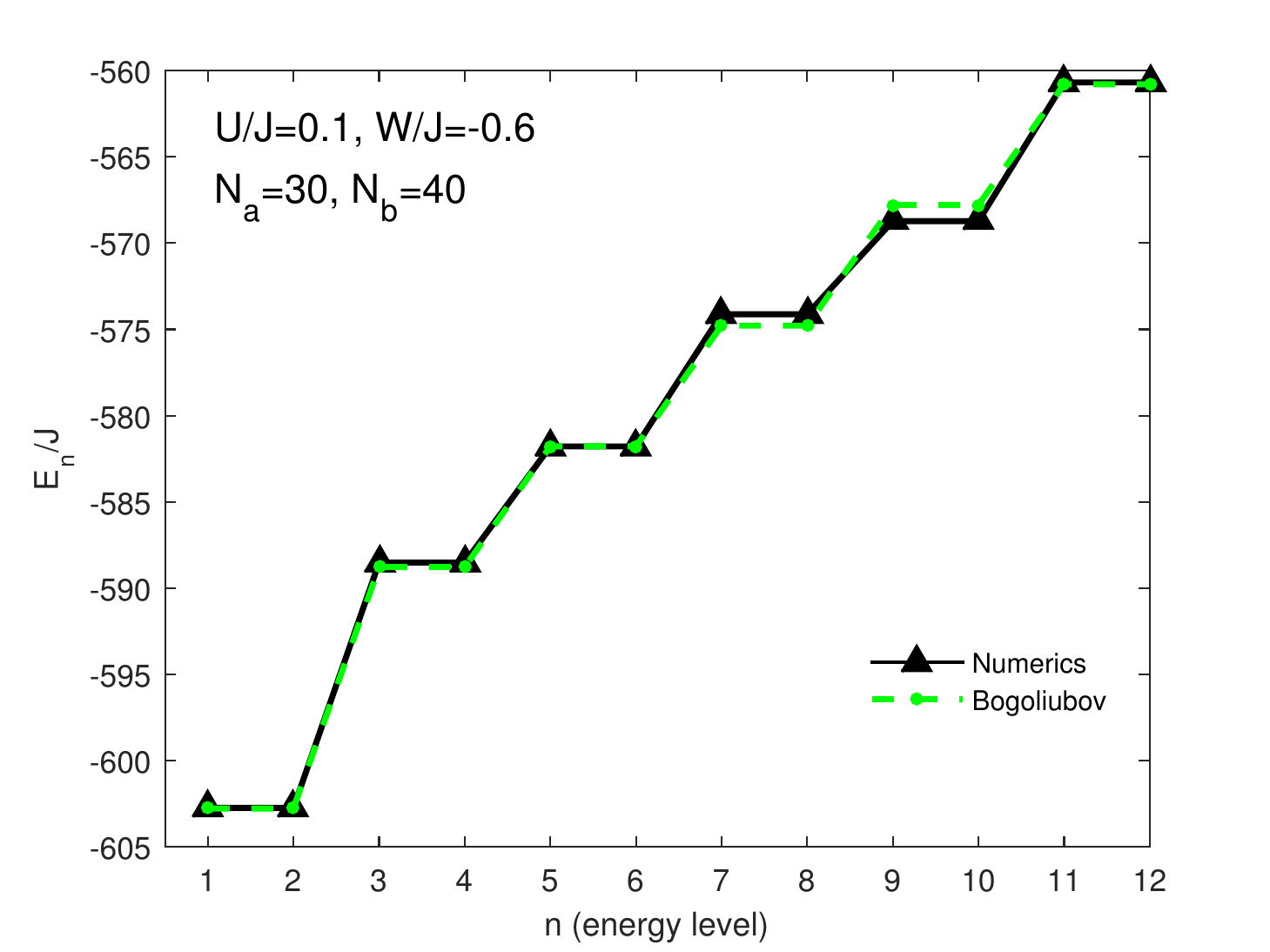}
\end{tabular}
\caption{(Color online) 
Case $W<0$ with $U_a=U_b \equiv U$.
Left panel: $|W|<U$. 
Black triangles: eigenvalues obtained numerically. 
Green dots: Eigenvalues obtained by Eq. (\ref{eigenvalueswsu}), 
$(N_\alpha, N_\beta)=(0,0),(1,0),(0,1),(1,1),(2,1),...$ correspond to the 
energy level $n=1,2,3,4,5,...$ ($n=1$ denotes the ground state). 
Right panel: $|W|>U$. 
Black triangles: eigenvalues obtained numerically. 
Green dots: Eigenvalues obtained by Eq. (\ref{eigenvalueswgua3}), 
$(N_{aR}, N_{bR})=(0,0),(0,1),(1,0),(1,1),(1,2),...$ correspond to the energy 
level $n=1,2,3,4,5,...$. 
All the quantities are dimensionless. Lines are for eye-guide.}
\label{fig7x}
\end{center}
\end{figure}
%

\begin{figure}[h]
\begin{center}
\begin{tabular}{c}
\includegraphics[
clip,width=0.6\textwidth ]{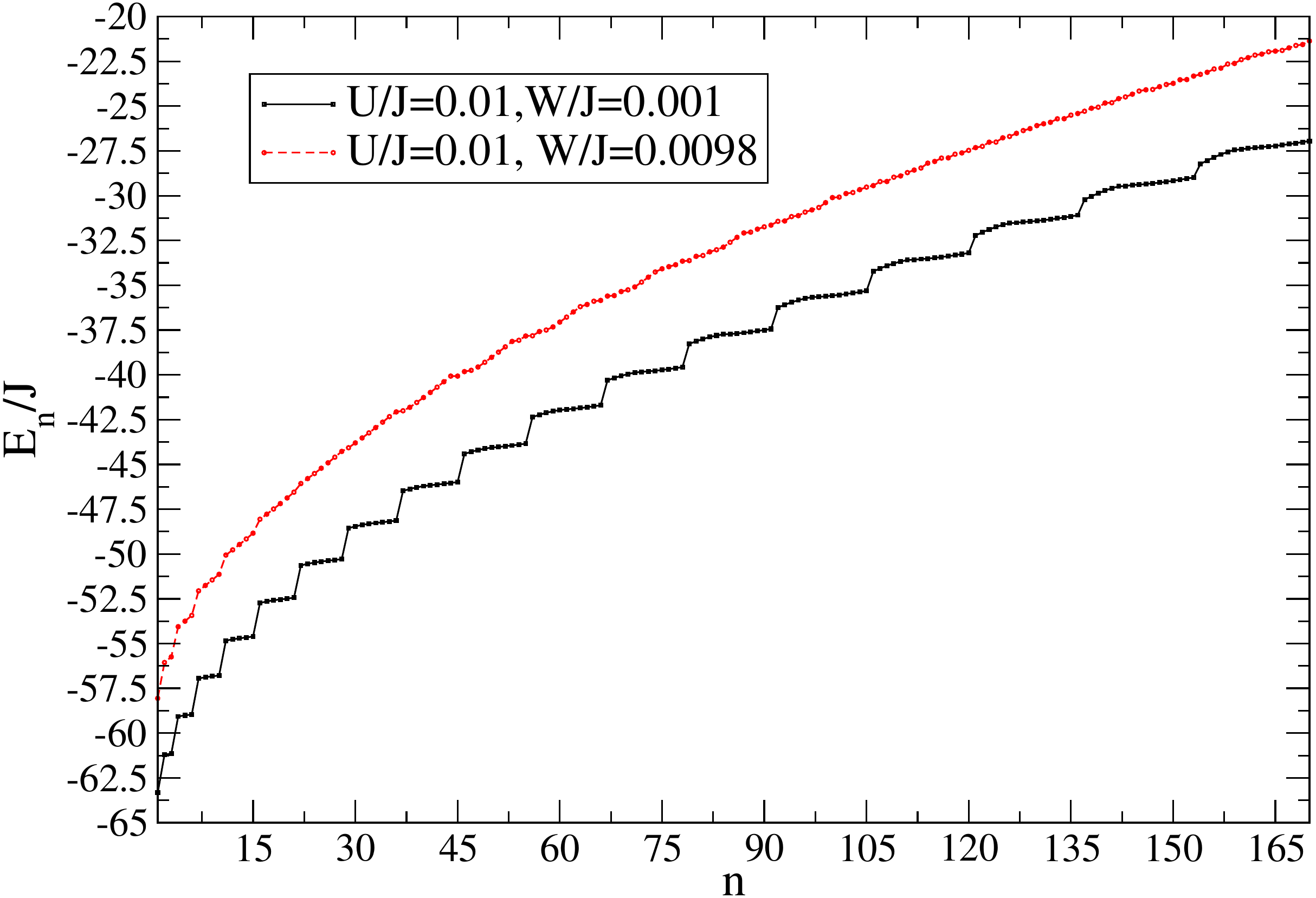}
\end{tabular}
\caption{(Color online) 
Numerical calculation of the energy levels of model
(\ref{bhtwo}). $N_a=30$, $N_b=40$,
$U_a/J=U_b/J \equiv U/J=0.01$. The spectral collapse is illustrated by comparing 
the spectrum for $W/J=0.001$ (squares-solid) and $W/J=0.0098$ (circle-dashed).
Horizontal axis: Excitation number $n$. Vertical axis: eigenvalues $E_n$ of
Hamiltonian (\ref{bhtwo}) in units of $J$.}
\label{contLIM}
\end{center}
\end{figure}

The left panels of Figures \ref{fig6x} and \ref{fig7x} corresponding to the repulsive 
and the attractive case, respectively, describe the eigenvalues of the 2-species Hamiltonian 
in the weak-interaction regime $|W| < U$. 
Such figures clearly show the excellent agreement between the eigenvalues provided by 
equation (\ref{eigenvalueswsu}) and the eigenvalues determined numerically.
By following the same ``Bogoliubov vs numerics" path, it is interesting to check if 
the expected transition to a continuous spectrum (spectral collapse)
discussed above, is observed.
To this end we have calculated the exact eigenvalues of model (\ref{bhtwo})
for two cases described in Fig. \ref{contLIM} by the squares-solid (red) line
($W << U$) and circles-dashed (black) line where the interspecies interaction $W \simeq U$.
The latter case represents the limit 
$U_aU_b-W^2 \to 0^+$ with $U_a=$ $ U_b=U$ (see formula (\ref{lim})). 
%
The comparison of the two lines clearly shows how the 
continuous character of the energy spectrum emerges in the limit $U -W\to 0^+$. 

The change $W>0 \to W <0$ does not affect the essence of the previous scheme so that the same 
conclusions can be found, in the attractive case,
when $|W|$ is increased and the demixing effect illustrated in Fig. \ref{fig3n} takes place.
The dramatic change of the algebraic structure of $\hat{H}$ reflecting the occurrence
of the spectral collapse suggests that a different mode picture must be used when $|W| > U$
in order to get the Hamiltonian into the diagonal form. This is discussed in the next subsection.
%
%
\subsection{Low-energy states with localized populations}
\label{52}

{\it Repulsive interaction}. For $W$ sufficiently larger than $U_c$
the boson distributions feature two possible forms
$$
\left
\{
\begin{array}{l l}
N_{aR} << {N}_{aL} \simeq N_a \, ,\quad & N_{bL} << {N}_{bR} \simeq N_b\, ,
\\
{\-}
\\
N_{bR} << {N}_{bL} \simeq N_b \, ,\quad & N_{aL} << {N}_{aR} \simeq N_a \, ,
\end{array}
\right .
$$
where two bosonic modes are macroscopically occupied and
${N}_{c} = {N}_{cL} +{N}_{cR}$ with $c= a,b$ are conserved quantities.
Such configurations are totally equivalent under the exchange, for each species,
of the left and right populations. The application of the Bogoliubov scheme
is discussed in \ref{due} for the case $N_{aR} << {N}_{a}$ and $N_{bL} << {N}_{b}$.
For this configuration the eigenvalues of $\hat{H}$ are found to be
\begin{equation}
\label{eigenvalueswgu1}
E_1({N}_{aR}, {N}_{bL}) \simeq E_0 ({N}_{a}, {N}_{b})
+ \sigma_a \; {N}_{aR} +  \sigma_b \; {N}_{bL} \; ,
\end{equation}
with
\begin{equation}
\label{e0}
E_0 ({N}_{a}, {N}_{b})
=
\sum_{c=a,b} \left[ \frac{U_c}{2} \bigl ({N}^2_{c} -{N}_{c} \bigr ) - \frac{J^2_c {N}_c}{\sigma_c} \right ]
%
%
\; ,
\end{equation}
and
\begin{equation}
\sigma_a = W {N}_{b} -U_a {N}_{a} \, ,\quad \sigma_b = W {N}_{a} -U_b {N}_{b}\; .
\label{sigma}
\end{equation}
Note that the general conditions $U_a \ne U_b$ and $J_a \ne J_b$ have been kept
since they do not affect the diagonalization process.
The eigenvalues of the 2-species Hamiltonian in the strong-interaction regime $|W| > U$
are illustrated in the right panel of Figure \ref{fig6x} in the 
repulsive case. 
Also in this case, the eigenvalues provided by 
equation (\ref{eigenvalueswgu1}) and the eigenvalues determined numerically
show an excellent agreement. The corresponding eigenstates are
\begin{equation}
|E_1({N}_{bL}, {N}_{aR})\rangle
\! = \! |{N}_{aL} \rangle |z_{bL}, {N}_{bL} \rangle |z_{aR}, {N}_{aR}\rangle |{N}_{bR} \rangle
\label{estate2}
\end{equation}
(recall that ${N}_{bR}=N_b-{N}_{bL}$ and ${N}_{aL} =$ $N_a -{N}_{aR}$) where
the two generalized Glauber coherent states \cite{solomon}
$$
|z_{bL}, {N}_{bL} \rangle = \hat{D}(z_{bL}) |{N}_{bL} \rangle
\; ,
\quad
|z_{aR}, {N}_{aR}\rangle =\hat{D}(z_{aR}) |{N}_{aR}\rangle
$$
incorporate the displacement operators $\hat{D}(z_{aR})$ and $\hat{D}(z_{bL})$.
The latter ensure the reduction of Hamiltonian $\hat{H}$ to the diagonal form for
$z_{aR} \equiv J_a \sqrt{N_a} /\sigma_a$, and $z_{bL} \equiv J_b \sqrt{N_b} / \sigma_b$.
Operators
$\hat{D}(z_{aR})$ and $\hat{D}(z_{bL})$ are defined in \ref{due}.
Formula (\ref{eigenvalueswgu1}) shows that the lowest energy state of the double-well system
is found for ${N}_{bL}={N}_{aR}=0$ which gives
$$
|E_1(0, 0)\rangle =
| {N}_{a} \rangle |z_{bL} \rangle |z_{aR} \rangle |{N}_{b} \rangle\, .
$$
In this case $|z_{bL},{N}_{bL} \rangle$ and $|z_{aR}, {N}_{aR}\rangle$ reduce
to two Glauber coherent states $|z_{bL} \rangle$ and $|z_{aR} \rangle$ implying that
the minimum-uncertainty relation of boson operators reaches its optimized form
in the ground state (see \ref{due}).

The same diagonalization scheme can be applied to the case
${N}_{aL} \ll {N}_{aR}$, ${N}_{bR}\ll {N}_{bL}$ and implies the approximation $H \simeq H_2$
described by formula (\ref{Hlin2}). The eigenvalues and the eigenstates
of $\hat{H}_2$ are found to be
\begin{equation}
\label{eigenvalueswgu2}
E_2({N}_{aL}, {N}_{bR}) \simeq E_0 ({N}_{a}, {N}_{b})
+ \sigma_a \; {N}_{aL} +  \sigma_b \; {N}_{bR} \; ,
\end{equation}
$$
|E_2({N}_{aL}, {N}_{bR})\rangle
=
|z_{aL}, {N}_{aL} \rangle | {N}_{bL} \rangle |{N}_{aR} \rangle |z_{bR}, {N}_{bR}\rangle\; ,
$$
and, following the same lines leading to equations (\ref{z1}), one determines the 
conditions
$z_{aL} \equiv J_a \sqrt{N_a} /\sigma_a$, and $z_{bR} \equiv J_b \sqrt{N_b} / \sigma_b$
taking the Hamiltonian into the diagonal form.
For ${N}_{aL} =$ $ {N}_{bR}=0$ the lowest energy eigenvalue is achieved which
satisfies $E_2(0, 0)= E_0 ({N}_{a}, {N}_{b})\equiv E_1(0, 0)$ and is associated with
$|E_2(0, 0)\rangle = |z_{aL} \rangle | N_a \rangle |N_b \rangle |z_{bR}\rangle$
again containing two Glauber coherent states.
Due to the degeneracy of the Bogoliubov spectrum
(eigenvalues (\ref{eigenvalueswgu1}) and (\ref{eigenvalueswgu2}) are identical),
the obvious form of the ground state (gs) is thus given by
$$
|E_{gs} \rangle = \frac{1}{\sqrt 2} \Bigl ( |E_1(0, 0)\rangle +  |E_2 (0, 0)\rangle \Bigr )\, ,
$$
%
which well reproduces qualitatively the behavior of the ground state illustrated in the second row
of Fig. \ref{fig2n} for $W$ sufficiently larger than $U$.
Weakly-excited states can be derived together with their eigenvalues
by constructing suitable symmetrized combinations of equal-energy states
$|E_\pm (q, p)\rangle = (|E_1(q, p)\rangle \pm |E_2(p, q)\rangle)/\sqrt 2$
where ${N}_{aR}=$ ${N}_{aL}= q$ and ${N}_{bR}= {N}_{bL}= p$.
The degeneracy characterizing such states, well known for a system of two symmetric wells with
a single atomic species \cite{milb}, can be removed by applying to states $|E_\pm (q, p)\rangle$ the
approximation scheme developed in reference \cite{LaLi} for a double-well potential.

\noindent
{\it Attractive interaction}. In the strong-interaction regime, namely for $|W|$ sufficiently 
larger than $U$, the two configurations
$$
\left
\{
\begin{array}{l l}
N_{aR} << {N}_{aL} \simeq N_a \, ,\quad & N_{bR} << {N}_{bL} \simeq N_b\, ,
\\
{\-}
\\
N_{aL} << {N}_{aR} \simeq N_a \, ,\quad & N_{bL} << {N}_{bR} \simeq N_b\, ,
\end{array}
\right .
$$
--both entailing the localization of the two species in one of the two wells--
describe the ground state. Due to the attractive interaction the macroscopically
occupied modes are those corresponding to the same well for each population. The
diagonal Hamiltonian of this case features the eigenvalues 
\begin{equation}
\label{eigenvalueswgua1T}
E_3({N}_{aR}, {N}_{bR}) \simeq {\tilde E}_0 ({N}_{a}, {N}_{b})
+ \sigma_a \; {N}_{aR} +  \sigma_b \; {N}_{aL} \; ,
\end{equation}
derived in \ref{tre}. 
The equivalent spectrum found by exchanging left and right populations
is given by formula (\ref{eigenvalueswgua4}).  
The right panel of Figure \ref{fig7x} illustrates the Bogoliubov spectrum
(\ref{eigenvalueswgua1T}) and the numerical
spectrum for the TDS model in the strong-interaction attractive case
$|W| > U$. 
As in the repulsive case, an excellent agreement characterizes, at low energies, 
the Bogoliubov and numerical eigenvalues.

\begin{figure}[h]
\begin{center}
\begin{tabular}{cc}
\includegraphics[
clip,width=0.5\textwidth ]{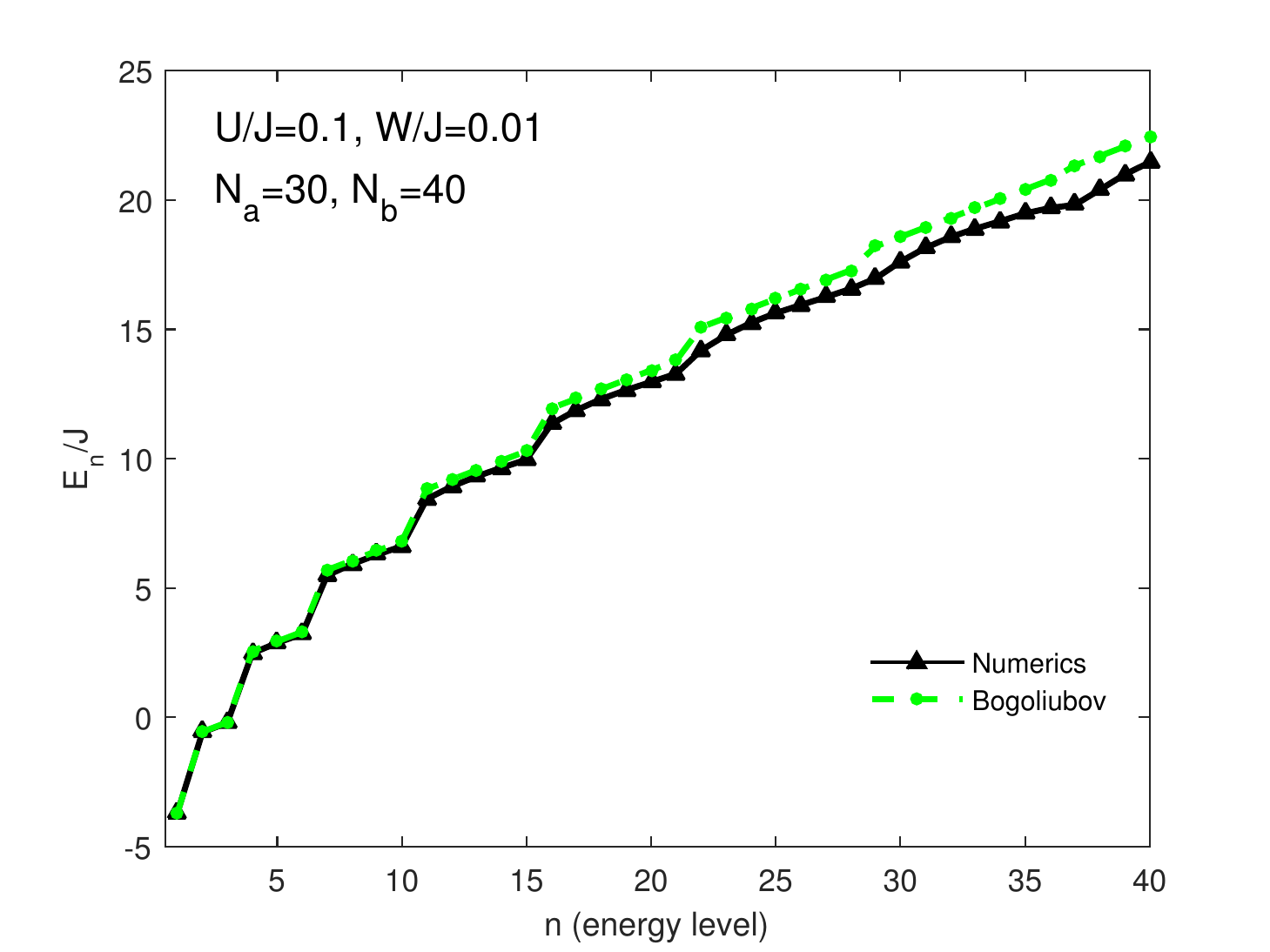}
&
\includegraphics[
clip,width=0.5\textwidth ]{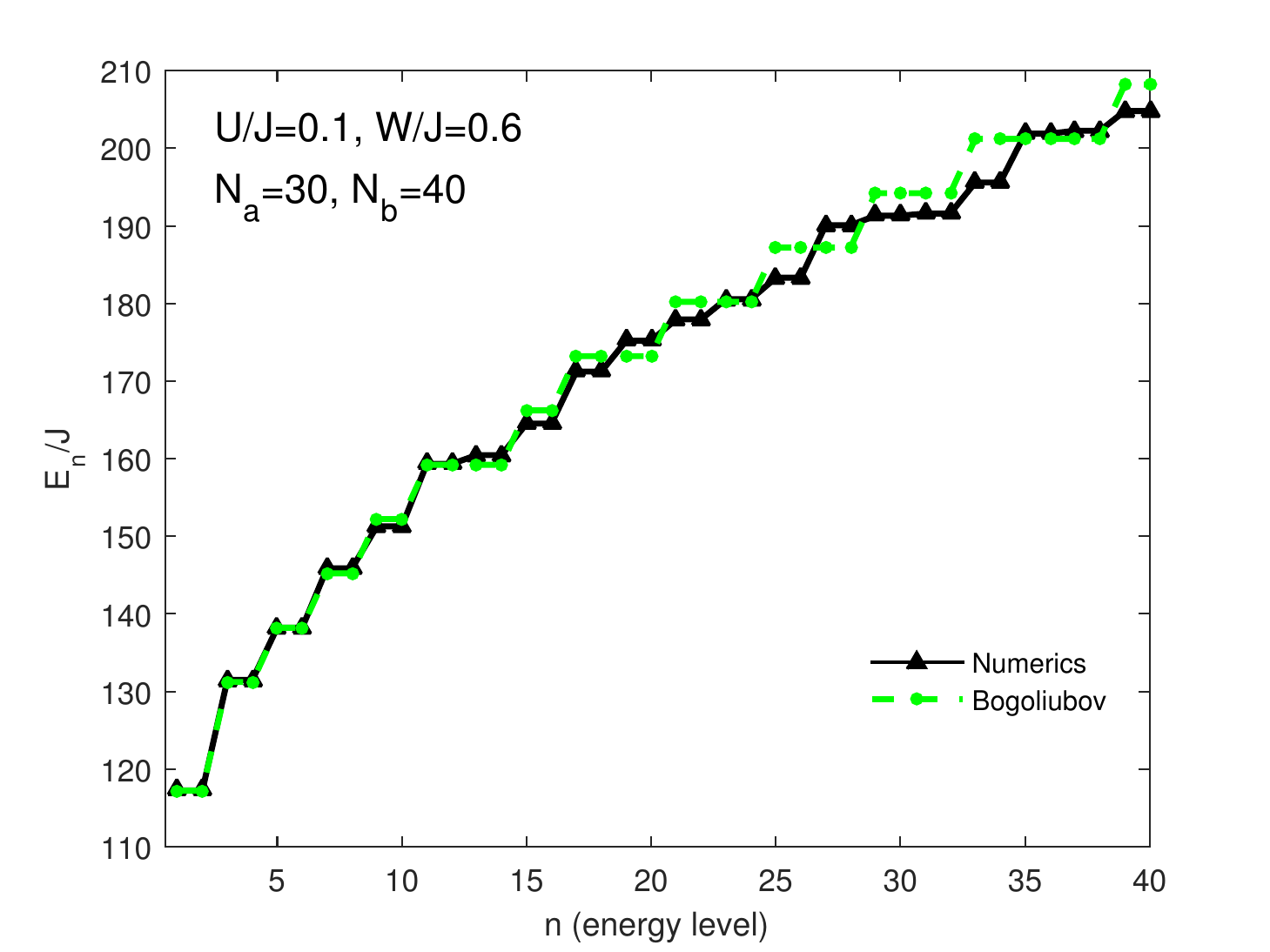}
\end{tabular}
\caption{(Color online) 
In these panels we show that the agreement of the Bogoliubov and the
numerical spectrum remains satisfactory even for a larger number of energy levels. 
Deviations become more and more visible for $n$ larger than $n \sim 30$.  
Both the case $W<U$  (left panel) and $W>U$ (right panel) are considered for $W>0$. }
\label{mlevels}
\end{center}
\end{figure}

\begin{figure}[h]
\begin{center}
\begin{tabular}{c}
\includegraphics[
clip,width=0.7\textwidth ]{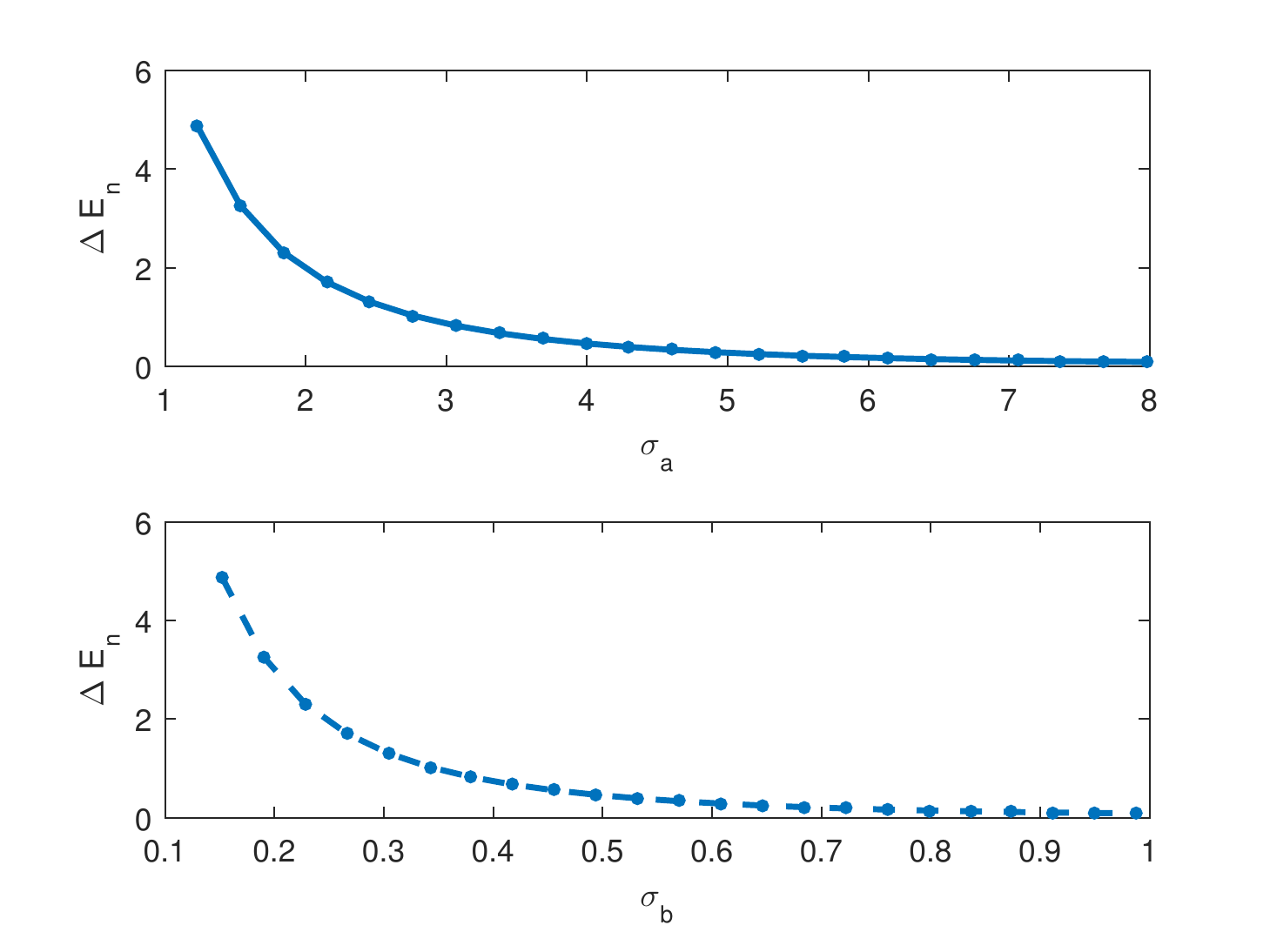}
\end{tabular}
\caption{(Color online) Relative error $\Delta E_n$, described by formula (\ref{re}), as a function 
of $\sigma_a$ (upper panel) and $\sigma_b$ (lower panel), for $W= 0.169$, and $U=0.1$ 
(see eq. (\ref{sigma})). Both $\sigma_a$ and 
$\sigma_b$ are varied increasing $N$ with $N_a/N_b= 0.66$.  
The two panels show how the Bogoliubov spectrum approaches the numerical spectrum in the
limit $N \gg 1$.}
\label{dev}
\end{center}
\end{figure}

{\it Concluding remarks}. To better illustrate the range of validity of the Bogoliubov 
scheme, we compare the Bogoliubov and the numerical spectrum for a number of energy 
levels ($n= 40$) larger than that of Figs. \ref{fig6x} and \ref{fig7x}. 
Fig. \ref{mlevels} shows that their agreement remains qualitatively satisfactory 
when $n$ is increased. 
Both the case $W<U$  (left panel) and $W>U$ (right panel) are considered for $W>0$.
The same results can be shown to hold in the attractive case. 
For $|W| > U$ and large number of bosons, the agreement between the exact numerical diagonalization and the Bogoliubov scheme can be estimated by calculating 
the relative error as a function of parameters $\sigma_a$ and $\sigma_b$. 
This is defined by 
\begin{equation}
\Delta E_{n}
= \left[  \frac{\sum_n (E_{n}-E_{n,B})^2}{\sum_n E_{n}^2} \right ]^{1/2}
\; ,
\label{re}
\end{equation}
where $E_{n}$ and $E_{n,B}$ refer to the levels obtained numerically and from
the Bogoliubov spectrum (\ref{eigenvalueswgu1}), 
respectively. Fig. \ref{dev} shows the dependence of $\Delta E_{n}$ from 
$\sigma_a$ and $\sigma_b$ at fixed $W>0$ and $U$. 
$\sigma_a$ and $\sigma_b$ can be varied by increasing the boson number 
$N$ since $\sigma_a = (3W-2U)N/5$ and $\sigma_b= (2W-3U)N/5$ 
with a fixed population ratio $N_a/N_b = 2/3$ (see (\ref{sigma})). 
The rapid decreasing of $\Delta E_{n}$ for large $N$ (in Fig. \ref{dev},
$\sigma_a =8$, $\sigma_b = 1$ correspond to $N=130$) 
highlights that, in the semiclassical limit, 
the Bogoliubov spectrum approaches the numerical one
(the same results are found in the attractive case $W<0$).
The only {\it caveat} in applying the Bogoliubov scheme 
concerns the ratio $N_a/N_b$ and $U/W$. A careful choice of the latter quantities must be 
done in order to avoid possible diverging behavior of the constant energy (\ref{e0}).
This pathology does not affect the Bogoliubov spectrum found in section (\ref{51}) 
for $|W| < U$.
%

\section{Conclusions}
We have studied a bosonic binary mixture confined by
a one-dimensional double-well potential (TSD model) where each atomic species
is described by a BH model and the two species are coupled via an onsite
density-density interaction $W$.
We have investigated the system by adopting two different approaches and analyzed 
the ground state by varying $W$ in both the repulsive ($W>0$) and the attractive 
($W<0$) case.

First, we have diagonalized numerically the TSD model and obtained
an exact description of the ground-state structure. Then, we have characterized
the TSD lowest energy state from the viewpoint of correlation properties.
In this respect, we have studied (the appropriate specializations to the case under 
study of) the quantum Fisher information, the coherence visibility and the 
entanglement entropy as functions of $W$, showing their sensitivity to the
changes of the ground-state structure.

As a second, after identifying the weakly-occupied bosonic modes
in the four regimes characterized by $W$ (interspecies interaction) and 
$U_c$ (onsite interaction of species $c=a,b$), we have reconstructed within the Bogoliubov approach 
the energy spectrum and the relevant low-energy states.
The low-energy part of the spectrum displays an excellent agreement with the spectrum 
calculated numerically. This agreement is shown to improve when the
total boson number $N$ is increased.
Finally, we have shown how moving from the delocalized regime to the localized one 
--where both species feature macroscopic components, separated ($W>0$) or mixed ($W<0$)-- determines a spectral collapse in which the interlevel distance became smaller and smaller 
when $W^2$ approaches  $U_a U_b$ from below.
The eigenstates, exhibiting squeezed-state components for $U_a U_b > W^2$,
assume a form characterized by Glauber-state components for $W^2 > U_a U_b$,
reflecting the new algebraic structure of the Hamiltonian in this regime.


\section*{Aknowledgments} 
The authors acknowledge the Ministero dell'Istruzione, Universit\`a e Ricerca,
(MIUR) for the support provided by the Grant PRIN 2010LLKJBX.
%

\appendix

\section{Reduction of $\hat{H}$ to a diagonal form in the regime $|W| < U_c$}
\label{uno}

Within the Bogoliubov scheme, operators  $\hat{A}$, $\hat{B}$ can be seen as complex parameters
if $N_a \gg N_\alpha$ and $N_b \gg N_\beta$. With the substitutions
$\hat{A}\to e^{i\phi_A}\sqrt {N_a} $, $\hat{B} \to  e^{i\phi_B} \sqrt {N_b}$ (in which
the phase factors associated to $\hat{A}$ and $\hat{B}$ can be removed due to the U(1) symmetry of the Hamiltonian)
model (\ref{bhtwo}) reduces to
%
\begin{eqnarray}
&&
\hat{H}
=
\frac{W}{2} \sqrt { {N}_a {N}_b} \; ( \hat{\alpha}  + \hat{\alpha}^\dagger )\; (\hat{\beta}  + \hat{\beta}^\dagger )
+ 2J_a {\hat N}_\alpha  + 2J_b {\hat N}_\beta
\nonumber\\
&&
+\frac{U_a}{4} {N}_a (\hat{\alpha}  + \hat{\alpha}^{\dagger} )^2
+ \frac{U_b}{4} {N}_b (\hat{\beta}  + \hat{\beta}^{\dagger } )^2
+K ,
\label{albe}
\end{eqnarray}
where the terms depending on $N_a$, $N_b$ are grouped in the constant
$$
K =
\frac{U_a}{4}({N}^2_a -{N}_a) +\frac{U_b}{4}({N}^2_b -{N}_b)
+ \frac{W}{2} {N}_a {N}_b - JN\; ,
$$
and third-order (or higher-order) terms have been neglected. We focus our attention
on the special case $J_a = J_b \equiv J$ in which the technicalities of the diagonalization
scheme are not excessive and the ground-state properties are easily evidenced.
The general case $J_a \ne J_b$ will be discussed elsewhere.
The diagonalization process is effected in two steps.
First, by exploiting the unitary transformation $\hat{U} = \exp (i\phi \hat{L} )$
with $\hat{L} = i ( \hat{\alpha} \hat{\beta}^\dagger - \hat{\beta} \hat{\alpha}^\dagger )$,
which generates the mode transformations
$$
\left
\{
\begin{array}{ll}
\hat{U}^\dagger (\hat{\alpha} + \hat{\alpha}^\dagger)\; \hat{U} = (\hat{\alpha} + \hat{\alpha}^\dagger) \cos(\phi)+ (\hat{\beta} + \hat{\beta}^\dagger) \sin(\phi)
\, ,
\\
{\-}
\\
\hat{U}^\dagger (\hat{\beta} + \hat{\beta}^\dagger)\; \hat{U} =  (\hat{\beta} + \hat{\beta}^\dagger) \cos(\phi)-(\hat{\alpha} + \hat{\alpha}^\dagger) \sin(\phi)
\, .
\end{array}
\right .
$$
For $\phi$ such that ${\rm tg} (2\phi) = 2W \sqrt{N_aN_b}/(U_bN_b-U_aN_a )$ (obtained by requiring 
that the term coupling modes $\alpha$ and $\beta$ vanishes in the new Hamiltonian),
these transformations supply the mode-decoupled formula $\hat{H}_{dec} = \hat{U} \hat{H} \hat{U}^\dagger$, 
whose explicit expression is
%
\begin{eqnarray}
&&
\hat{H}_{dec} = K-2J+ \left (J+ \frac{u-D}{8}\right ) (2N_\alpha +1)
+ \frac{u-D}{8}\; ( \alpha^{\dagger 2}+ \alpha^{ 2} )
\nonumber\\
&&
+\left (J+ \frac{u+D}{8}\right )  (2N_\beta +1) 
+ \frac{u+D}{8} \; ( \beta^{\dagger 2}+ \beta^{ 2} )
\; ,
\label{Dalbe}
\end{eqnarray}
in which
$$
D =  \sqrt{ \Delta^2 +4W^2N_aN_b }\; , \quad
u = U_aN_a +U_bN_b\; , \quad \Delta = U_aN_a -U_bN_b\; .
$$
We have tacitly assumed $U_bN_b >U_aN_a$ in ${\rm tg} (2\phi)$.
Note that, if $U_bN_b <U_aN_a$, the angle $\phi$ will be negative and
the definitions of $R_\alpha$, $R_\beta$ simply exchange
with each other.

The second step follows from the observation that
the two subHamiltonians for modes $\alpha$ and $\beta$ are linear combinations of
the generators of the two algebras SU(1,1) $\hat{K}^\sigma_{3}$ $= (2\hat{\sigma}^{\dagger} \hat{\sigma} +1)/4$,
$\hat{K}^\sigma_+ = \hat{\sigma}^{\dagger 2}/2$ and $\hat{K}^\sigma_- =(\hat{K}^\sigma_+)^\dagger$
where $\hat{\sigma}= \hat{\alpha} , \hat{\beta}$. Then they can be reduced to the diagonal form
(see, for example, Ref. \cite{cavpe}) by using the simple SU(1,1)-group transformations
$\hat{S}_\sigma = \exp[\theta_\sigma (\hat{K}^\sigma_+ - \hat{K}^\sigma_-)/2]$ having the typical form
of a squeezing operator.
%
The diagonal Hamiltonian is given by
$$
\hat{H}_D 
=\hat{S}^\dagger_\alpha \hat{S}^\dagger_\beta \hat{H}_{dec} \hat{S}_\beta \hat{S}_\alpha
= K-2J + \sqrt {R_\alpha } \; (2{\hat N}_\alpha+1)+ \sqrt {R_\beta } \; (2{\hat N}_\beta+1)
\; ,
$$
with $R_{\alpha} = J[ J+(u-  D)/4]$, $R_{\beta} = J[ J+(u + D)/4]$, when
the angles $\theta_\alpha$ and $\theta_\beta$ are assumed to satisfy the condition 
${\rm th} (2 \theta_\sigma) = (R_\sigma -J^2)/(R_\sigma +J^2)$ with $\sigma= \alpha, \beta$.
%

\section{Reduction of $\hat{H}$ to a diagonal form in the regime $|W| > U_c$}
\label{due}
Due to $N_{aR} << {N}_{a}$ and $N_{bL} << {N}_{b}$,
the Bogoliubov approximation reduces $\hat{H} $ to the effective Hamiltonian involving
modes $\hat{a}_{R}$ and $\hat{b}_{L}$
\begin{equation}
\hat{H}_1 = C(N_a, N_b)
+ \sigma_a  {\hat N}_{aR} +  \sigma_b {\hat N}_{bL}
-J_a \sqrt {{N}_{a}} \; (\hat{a}_{R}^{\dagger}+ \hat{a}_{R})
-J_b \sqrt {{N}_{b}} \; (\hat{b}_{L}^{\dagger}+ \hat{b}_{L})
\; ,
\label{Hlin1}
\end{equation}
with
$$
C(N_a, N_b) = \frac{1}{2} \sum_{c=a,b} U_c \Bigl ({N}^2_{c} -{N}_{c} \Bigr )
\; ,\quad
\sigma_a = W {N}_{b} -U_a {N}_{a} \; ,\quad \sigma_b = W {N}_{a} -U_b {N}_{b}
\; .
$$
The opposite case $N_{aL} << {N}_{a}$ and $N_{bR} << {N}_{b}$ leads to
the effective Hamiltonian  
%
\begin{equation}
\hat{H}_2
= C(N_a, N_b) + \sigma_a  {\hat N}_{aL} +  \sigma_b {\hat N}_{bR}
-J_a \sqrt {{N}_{a}} \; (\hat{a}_{L}^{\dagger}+ \hat{a}_{L})
-J_b \sqrt {{N}_{b}} \; (\hat{b}_{R}^{\dagger}+ \hat{b}_{R})\; .
\label{Hlin2}
\end{equation}
The eigenstates (\ref{estate2}) of $\hat{H_1}$ include
states $|z_{bL}, {N}_{bL} \rangle =$ $\hat{D}(z_{bL}) |{N}_{bL} \rangle$ and
$|z_{aR}, {N}_{aR}\rangle $$= \hat{D}(z_{aR}) |{N}_{aR}\rangle$ representing generalized
Glauber coherent states.
The two unitary transformations (taking $\hat{H}$ into the diagonal form)
$$
\hat{D}(z_{aR})= e^{z_{aR} \hat{a}_{R}^{\dagger} - z^*_{aR} \hat{a}_{R} }
\, ,\quad
\hat{D}(z_{bL}) =e^{z_{bL} \hat{b}_{L}^{\dagger}- z^*_{bL} \hat{b}_{L} }\, ,
$$
where
\begin{equation}
z_{aR} = J_a \sqrt{N_a} / \sigma_a\, ,\quad z_{bL} = J_b \sqrt{N_b} / \sigma_b\, ,
\label{z1}
\end{equation}
represent displacement operators of the Weyl-Heisenberg groups
associated to modes $\hat{a}_R$ and $\hat{b}_L$.
%
%
Here, the notation introduced in equation (\ref{superposition}) for a generic
Fock state $|i, j\rangle_L |N_a-i,N_b-j\rangle_R$ has been replaced by
the more effective fully-factorized one:
$|{N}_{aL}\rangle |{N}_{bL} \rangle |{N}_{aR}\rangle |{N}_{bR} \rangle$
where ${N}_{aL} = i$, ${N}_{bL}= j$, ${N}_{aR} = N_a -i$ and ${N}_{bR} = N_b -j$.
Similar to eigenstates (\ref{estate2}),
the states associated to eigenvalues (\ref{eigenvalueswgu2})
$$ |E_2({N}_{aL}, {N}_{bR})\rangle
= |z_{aL}, {N}_{aL} \rangle | {N}_{bL} \rangle |{N}_{aR} \rangle |z_{bR}, {N}_{bR}\rangle,
$$
with ${N}_{aR}= {N}_{a}-{N}_{aL}$, ${N}_{bL}= {N}_{b}-{N}_{bR}$,
include two generalized Glauber coherent states defined by
$|z_{aL}, {N}_{aL} \rangle = $ $\hat{D}(z_{aL})|{N}_{aL}\rangle$
and
$|z_{bR}, {N}_{bR}\rangle= $ $\hat{D}(z_{bR}) |{N}_{bR} \rangle$
where $\hat{D}$'s represent displacement operators.

Concerning the generalized Glauber state defined by
$|z, n \rangle = \hat{D}(z) |n \rangle$,
it is worth recalling that their distinctive feature consists in providing
the minimum-uncertainty relation $\Delta^2_x \Delta^2_p = (2n+1)^2/4$, where
the canonical operators $\hat{x}$ and $\hat{p}$ are related to the boson mode
$\hat{a}= (\hat{x}+i\hat{p})/\sqrt 2$, $|n\rangle$ is a number-operator state,
and $\hat{D}(z)= \exp (z\hat{a}^\dagger -z^*\hat{a})$.


\section{Diagonal Hamiltonian for $W<0$}
\label{tre}
In the attractive case, the Hamiltonians obtained from the Bogoliubov scheme entails the excited states
$$
|E_3({N}_{aR}, {N}_{bR})\rangle =
%
|{N}_{aL} \rangle |{N}_{bL} \rangle |z_{aR}, {N}_{aR}\rangle |z_{bR}, {N}_{bR} \rangle \, ,
$$
with ${N}_{aL}= N_a -{N}_{aR}$, ${N}_{bL}= N_b -{N}_{bR}$, and
$$
|E_4({N}_{aL}, {N}_{bL})\rangle =
%
|z_{aL}, {N}_{aL}\rangle |z_{bL}, {N}_{bL} \rangle  |{N}_{aR} \rangle |{N}_{bR} \rangle \, ,
$$
with ${N}_{aR}= N_a -{N}_{aL}$, ${N}_{bR}= N_b -{N}_{bL}$.
These provide the lowest-energy state (characterized by the absolute minimum uncertainty)
for ${N}_{aR}, {N}_{bR} = 0$ and ${N}_{aL}, {N}_{bL} = 0$, respectively. The
corrisponding eigenvalues are easily found to be
\begin{equation}
\label{eigenvalueswgua3}
E_3({N}_{aR}, {N}_{bR}) \simeq \tilde E_0 ({N}_{a}, {N}_{b})
+ \sigma_a \; {N}_{aR} +  \sigma_b \; {N}_{bR} \; ,
\end{equation}
\begin{equation}
\label{eigenvalueswgua4}
E_4({N}_{aL}, {N}_{bL}) \simeq \tilde E_0 ({N}_{a}, {N}_{b})
+ \sigma_a \; {N}_{aL} +  \sigma_b \; {N}_{bL} \; ,
\end{equation}
where
$\tilde E_0 ({N}_{a}, {N}_{b})=E_0 ({N}_{a}, {N}_{b})-|W|N_aN_b$
($E_0$ is given by Eq. (\ref{e0})) and the interspecies interaction $W$ has been
replaced with $|W|$ in $\sigma_a$ and $\sigma_b$.
As for eigenvalues (\ref{eigenvalueswgu1}) and (\ref{eigenvalueswgu2}),
formulas (\ref{eigenvalueswgua3}) and (\ref{eigenvalueswgua4}) describe
the same spectrum.


\section*{References}
\bibliographystyle{iopart-num}



\end{document}